\documentclass[12pt]{article}
\usepackage[dvips]{graphicx}
\usepackage{epsfig,cite}
\usepackage {amsmath}
\usepackage{color}
\usepackage{amssymb}
\usepackage{slashed}
\include{epsf}

\newcount\timecount
\newcount\hours \newcount\minutes  \newcount\temp \newcount\pmhours
\hours = \time
\divide\hours by 60
\temp = \hours
\multiply\temp by 60
\minutes = \time
\advance\minutes by -\temp
\def\hour{\the\hours}
\def\minute{\ifnum\minutes<10 0\the\minutes
            \else\the\minutes\fi}
\def\clock{
\ifnum\hours=0 12:\minute\ AM
\else\ifnum\hours<12 \hour:\minute\ AM
      \else\ifnum\hours=12 12:\minute\ PM
            \else\ifnum\hours>12
                 \pmhours=\hours
                 \advance\pmhours by -12
                 \the\pmhours:\minute\ PM
                 \fi
            \fi
      \fi
\fi
}

\def\monthname{\relax\ifcase\month 0/\or January\or February\or
   March\or April\or May\or June\or July\or August\or September\or
   October\or November\or December\else\number\month/\fi}

\def\bold#1{\setbox0=\hbox{$#1$}%
     \kern-.025em\copy0\kern-\wd0
     \kern.05em\copy0\kern-\wd0
     \kern-.025em\raise.0433em\box0 }

\def\beq{\begin{equation}}
\def\eeq{\end{equation}}

\def\ga{\mathrel{\raise.3ex\hbox{$>$\kern-.75em\lower1ex\hbox{$\sim$}}}}
\def\la{\mathrel{\raise.3ex\hbox{$<$\kern-.75em\lower1ex\hbox{$\sim$}}}}
\def\gev{{\rm \, Ge\kern-0.125em V}}
\def\tev{{\rm \, Te\kern-0.125em V}}
\def\gyr{{\rm \, G\kern-0.125em yr}}

\def\gappeq{\mathrel{\rlap {\raise.5ex\hbox{$>$}}
{\lower.5ex\hbox{$\sim$}}}}
\def\lappeq{\mathrel{\rlap{\raise.5ex\hbox{$<$}}
{\lower.5ex\hbox{$\sim$}}}}
\def\Toprel#1\over#2{\mathrel{\mathop{#2}\limits^{#1}}}

\def\m12{m_{1\!/2}}

\def\bea{\begin{eqnarray}}
\def\eea{\end{eqnarray}}
\def\beqn{\begin{eqnarray}}
\def\eeqn{\end{eqnarray}}

\def\beqar{\begin{eqnarray}}
\def\eeqar{\end{eqnarray}}

\begin{document}

\begin{titlepage}
\pagestyle{empty}
\baselineskip=21pt
\rightline{KCL-PH-TH/2016-03, LCTS/2016-02}
\rightline{CERN-TH/2016-011, IPPP/16/02}
\vskip 0.8in
\begin{center}
{\large {\bf Search for Sphalerons in Proton-Proton Collisions}}

\end{center}
\begin{center}
\vskip 0.4in
 {\bf John~Ellis}$^{1,2}$
and {\bf Kazuki~Sakurai}$^3$
\vskip 0.1in
{\small {\it
$^1${Theoretical Particle Physics and Cosmology Group, Physics Department, \\
King's College London, London WC2R 2LS, UK}\\
\vskip 0.1in
$^2${Theoretical Physics Department, CERN, CH-1211 Geneva 23, Switzerland}\\
\vskip 0.1in
$^3${Institute for Particle Physics Phenomenology, 
Department of Physics, University of Durham, Science Laboratories, South Road, Durham, DH1 3LE, UK}\\
}}
\vskip 0.6in
{\bf Abstract}
\end{center}
\baselineskip=18pt \noindent
{\small
In a recent paper, Tye and Wong (TW) have argued that sphaleron-induced transitions
in high-energy proton-proton collisions should be enhanced compared to previous calculations,
based on a construction of a Bloch wave function in the periodic sphaleron potential
and the corresponding pass band structure. Here we convolute the calculations of TW
with parton distribution functions and simulations of final states to explore
the signatures of sphaleron transitions at the LHC and possible future
colliders. We calculate the increase of sphaleron transition rates in proton-proton collisions at 
centre-of-mass energies of 13/14/33/100 TeV for different sphaleron barrier heights, 
while recognising that the rates have large overall uncertainties. We use a simulation
to show that LHC searches for microscopic black holes should have good
efficiency for detecting sphaleron-induced final states, and discuss their experimental 
signatures and observability in Run~2 of the LHC and beyond.
We recast the early ATLAS Run-2 search for microscopic black holes to constrain 
the rate of sphaleron transitions at 13~TeV, deriving a significant limit
on the sphaleron transition rate for the nominal sphaleron barrier height of 9~TeV.}


\vfill
\leftline{
January 2016}
\end{titlepage}
\baselineskip=18pt
 
\section{Introduction}

Non-perturbative effects in the electroweak sector of the Standard Model are predicted to
violate baryon ($B$) and lepton ($L$) conservation, violating the combination $B + L$
while conserving $B - L$. The first example was provided by electroweak instantons~\cite{instantons}, which yield
$\Delta B = 3$ transitions that are suppressed to unobservable levels
by factors $\sim \exp (- 2 \pi/\alpha_W)$, where $\alpha_W = g_W^2/4 \pi$
is the SU(2) coupling strength. The second example was provided by sphalerons~\cite{Manton,KM},
which are classical solutions of the electroweak field equations that interpolate between
vacua with different values of the Chern-Simons number, providing a potential barrier $E_{\rm Sph}$ to
$\Delta B = 3$ transitions that is expected to be $\simeq 9$~TeV. It has been thought that experimental
observation of sphaleron-induced transitions would also be unobservable for the foreseeable future 
\cite{Mueller:1991fa, Espinosa:1989qn, Ringwald:1989ee, Zakharov:1990dj, Khlebnikov:1990ue, Porrati:1990rk, Khoze:1990bm, Rubakov:1996vz},
because $\Delta n = \pm 1$ transitions would be suppressed by $\exp ({\cal O}(-4 \pi/\alpha_W))$.

However, this longstanding consensus has been challenged in a bold recent paper~\cite{TW} by
S.-H.~Henry~Tye and Sam~S.~C.~Wong (TW), who argue that sphaleron-induced
transition rates could be much larger than had been estimated previously. They argue that
an essential element in calculating the rate of $\Delta n \ne 0$ transitions is the periodic
nature of the effective Chen-Simons potential, which should be taken into account by constructing the corresponding Bloch
wave function. Their approach leads to a band structure for transitions through the sphaleron barrier,
resulting in a reduced suppression at energies $ < E_{\rm Sph}$ that disappears entirely at
energies $\ge E_{\rm Sph}$. As stressed in~\cite{TW}, this remarkable claim raises the
possibility that sphaleron-induced transitions might be observable at the LHC and higher-energy
proton-proton colliders. The experimental observation of such transitions would not only be a
beautiful confirmation of profound theoretical insights, but would also have important
cosmological implications, since sphalerons are thought to have played an essential r\^ole
in generating the baryon asymmetry of the Universe~\cite{Rubakov:1996vz, Kuzmin:1985mm,Cohen:1993nk,Trodden:1998ym,Morrissey:2012db,Shaposhnikov:1987tw}.

In this paper we follow up the suggestion of TW by calculating the energy dependence
of the rates for sphaleron-induced transitions in proton-proton collisions, including the factors
arising from quark parton distribution functions, and use simulations of their possible final states
to study the possible signatures of such transitions. As stressed by TW,
there are inevitable uncertainties in calculations of the rates for sphaleron-induced transitions, notably
including the sphaleron barrier height $E_{\rm Sph}$, the coefficient inside the exponential
suppression, and any possible prefactor. That said, our calculations encourage us
to explore how the rate for sphaleron-induced transitions might
be constrained by experiments at the LHC, possibly during its Run~2 that has now started.
Accordingly, we simulate the final states of sphaleron-induced transitions, 
demonstrating that the searches for microscopic black holes that have already been designed
would have good acceptance for sphaleron-induced final states, which would also possess additional distinctive
signatures. As an illustration, we constrain sphaleron transition rates by recasting
the results of the ATLAS Run-2 search for microscopic black holes using $\sim 3$ fb$^{-1}$ of data recorded at
13~TeV in 2015~\cite{ATLAS}. We find that these data already exclude a pre-exponential transition rate factor of 
unity for the nominal sphaleron barrier height of 9~TeV .

\section{Theoretical Background}

It is argued in~\cite{TW} that sphaleron transitions can be modelled by a one-dimensional
Schr\"odinger equation of the form
\begin{equation}
\left( - \frac{1}{2 m} \frac{\partial^2}{\partial Q^2} + V(Q) \right) \Psi(Q) \; = \; E \Psi(Q) \, ,
\label{schrod}
\end{equation}
where $m$ is an effective ``mass" parameter for the Chern-Simons number $n$
whose value was first calculated in~\cite{TW}, $Q \equiv \mu/m_W$ where 
$\mu$ is defined implicitly by $n \pi = \mu - \sin(2\mu)/2$,
and the effective potential is taken from~\cite{Manton}:
\begin{equation}
V(Q) \; \simeq \; 4.75 \left(1.31 \sin^2 (Q m_W) + 0.60 \sin^4 (Q m_W) \right) \, {\rm TeV} \, .
\label{VQ}
\end{equation}
Two evaluations of $m$ were discussed in~\cite{TW}: one based on~\cite{Manton} that yielded the
estimate $m = 17.1$~TeV, and the other based on~\cite{AKY} that yielded the estimate $m = 22.5$~TeV.
The final results for the rate of sphaleron-induced transitions were very similar, and here we follow~\cite{TW}
in adopting the \cite{Manton}-based calculation that led to $m = 17.1$~TeV.

The sphaleron barrier height $E_{\rm Sph}$ is given by
\begin{equation}
E_{\rm Sph} \; = \; {\rm Max} \, V(Q) \; = \; V \left( \frac{\pi}{2 m_W} \right) \, .
\label{ESph}
\end{equation}
In a pure SU(2) theory, one finds $E_{\rm Sph} = 9.11$~TeV, and it is estimated that
incorporating the U(1) of the Standard Model reduces this by $\sim 1$\%. Here we follow~\cite{TW}
in assuming a nominal value of $E_{\rm Sph} = 9$~TeV, while presenting some numerical results for the alternative
choices $E_{\rm Sph} = 8, 10$~TeV. Later, we also use a recast of early Run-2 searches for microscopic
black holes to constrain the sphaleron transition rate as a function of $E_{\rm Sph}$.

As was discussed in detail in~\cite{TW}, the Bloch wave function for the periodic potential (\ref{VQ})
is straightforwardly obtained, and the corresponding conducting (pass) bands can be calculated, as well
as their widths and the gaps between the bands. The lowest-lying bands are very narrow, but the
widths increase with the heights of the bands. Averaging over the energies $E_{1,2}$ of the colliding quark partons
yields a strong suppression at $E_1 + E_2 \ll E_{\rm Sph}$, which corresponds to the
exponential suppression found in a conventional tunnelling calculation. However, this suppression
decreases as $E_1 + E_2 \to E_{\rm Sph}$, and there is no suppression for $E_1 + E_2 \ge E_{\rm Sph}$.

The result of the analysis in ~\cite{TW} can be summarized in the partonic cross-section 
\begin{equation}
\sigma ( \Delta n = \pm 1 ) \; \propto \; \exp \left( c \frac{4 \pi}{\alpha_W} S(E) \right) \, ,
\label{sigmaTW}
\end{equation}
where $E$ is the centre-of-mass energy of the parton-parton collision,
$c \sim 2$ and the suppression factor $S(E)$ is shown in Fig.~8 of~\cite{TW}. 
As seen there, it rises from the value $S(E) = - 1$ in the low-energy limit ($E \ll E_{\rm Sph}$) to $S(E) = 0$ for energies $E \ge E_{\rm Sph}$, with very similar results being found in~\cite{TW} for calculations
based on the work of~\cite{Manton} and~\cite{AKY}. 
For the purpose of our numerical calculations, we approximate $S(E)$ at intermediate energies by
\begin{equation}
S(E) \; = \; (1 - a) {\hat E} + a {\hat E}^2 - 1 \; \; \; \; {\rm for} \; \; \; \; 0 \; \le {\hat E} \; \le \; 1 \, ,
\label{numericalS}
\end{equation}
where ${\hat E} \equiv E/E_{\rm Sph}$ and $a = - 0.005$.

The overall magnitude of Eq.~\eqref{sigmaTW} is not given.
We speculate that the relevant scale should be proportional to 
the non-perturbative electro-weak cross-section for $q$-$q$ scattering, $\sigma_{qq}^{\rm EW}$.
Analogously to the fact that the inelastic $p$-$p$ cross-section is given roughly by $\sim 1/m_\pi^2$,
we take $\sigma_{qq}^{\rm EW} \sim 1/m_W^2$.  
Our cross-section formula is, thus, given as
\beqn
\sigma(\Delta n = \pm 1) 
&=& 
\frac{p}{m_W^2} \sum_{ab} \int d E \frac{d {\cal L}_{ab}}{d E} 
\exp \Big( c \frac{4 \pi}{\alpha_W} S( E ) \Big) \,,
\label{sigma}
\eeqn
where $p$ is an unknown constant and
$\frac{d {\cal L}_{ab}}{d E}$ is the parton luminosity function 
of the colliding quarks $a$ and $b$,
which are obtained from the parton distribution functions at a momentum fraction $x$, $f_a(x)$, evaluated at the appropriate energy scale $E$:
\beq
\frac{d {\cal L}_{ab}}{d E} = \frac{2E}{E_{\rm CM}^2} \int_{\ln \sqrt{\tau}}^{-\ln \sqrt{\tau}} d y f_a(\sqrt{\tau} e^y) f_b(\sqrt{\tau} e^{-y}),
\eeq
where $E_{\rm CM}$ is the centre-of-mass energy of the $p$-$p$ collision and $\tau = E^2 / E^2_{\rm CM}$.

\section{Cross-Section Calculations}

We include in our calculations collisions of all quarks and antiquarks in the lightest two
generations, namely $u, d, s$ and $c$. We recall that only left-handed (SU(2) doublet) quarks are active in inducing
sphaleron transitions, so that the usual unpolarized quark-quark parton collision luminosity functions must be reduced by a factor 4.
Additionally, we expect that quarks in the same generation must collide in an antitriplet state, reducing the corresponding
luminosity functions by another factor 3. In principle, one should also incorporate Cabibbo mixing, but this is unimportant
compared with the uncertainties in the calculation. 

\begin{figure}[t!]
\begin{center}
\includegraphics[height=10cm]{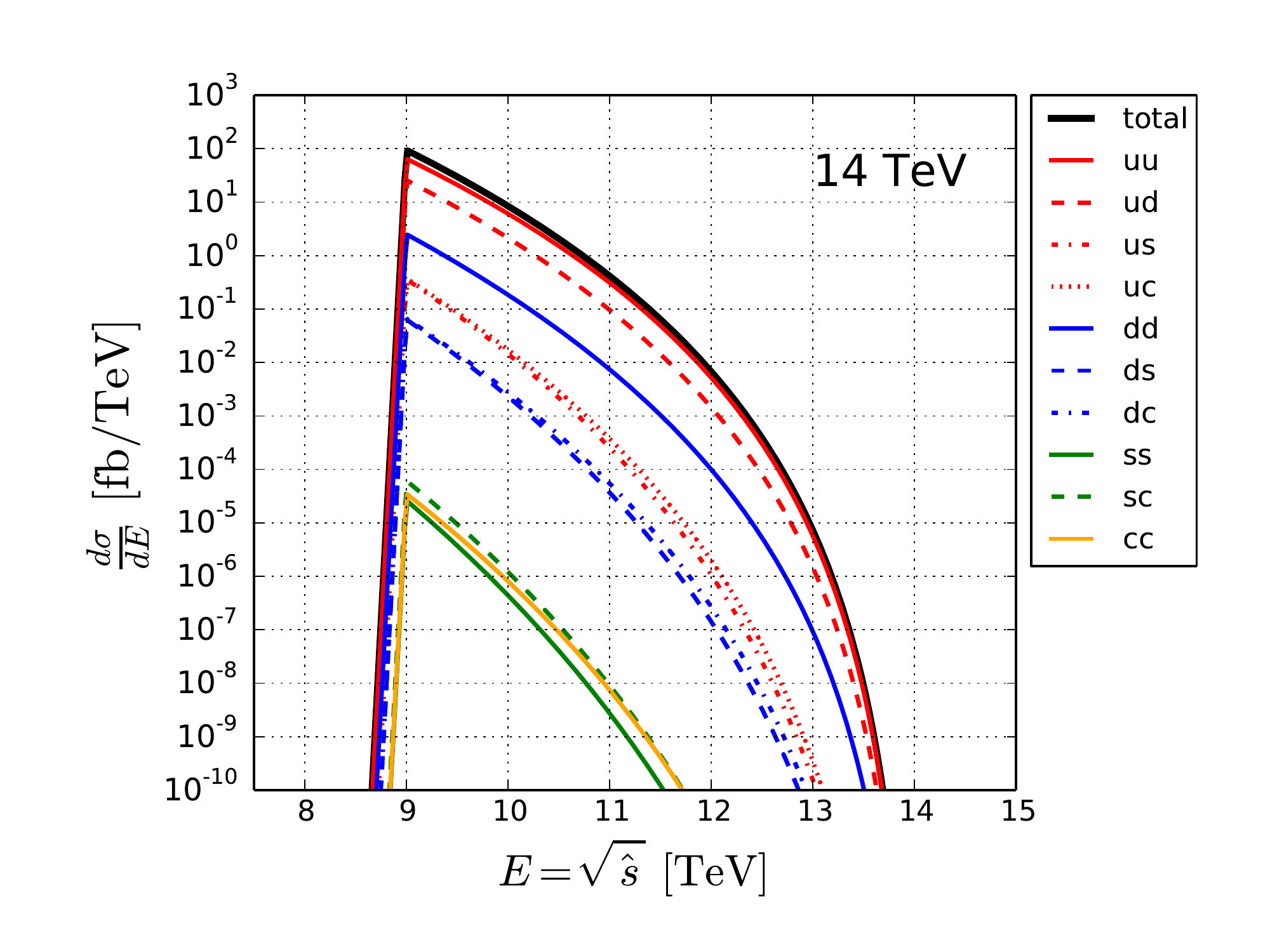} \\
\includegraphics[height=4.28cm]{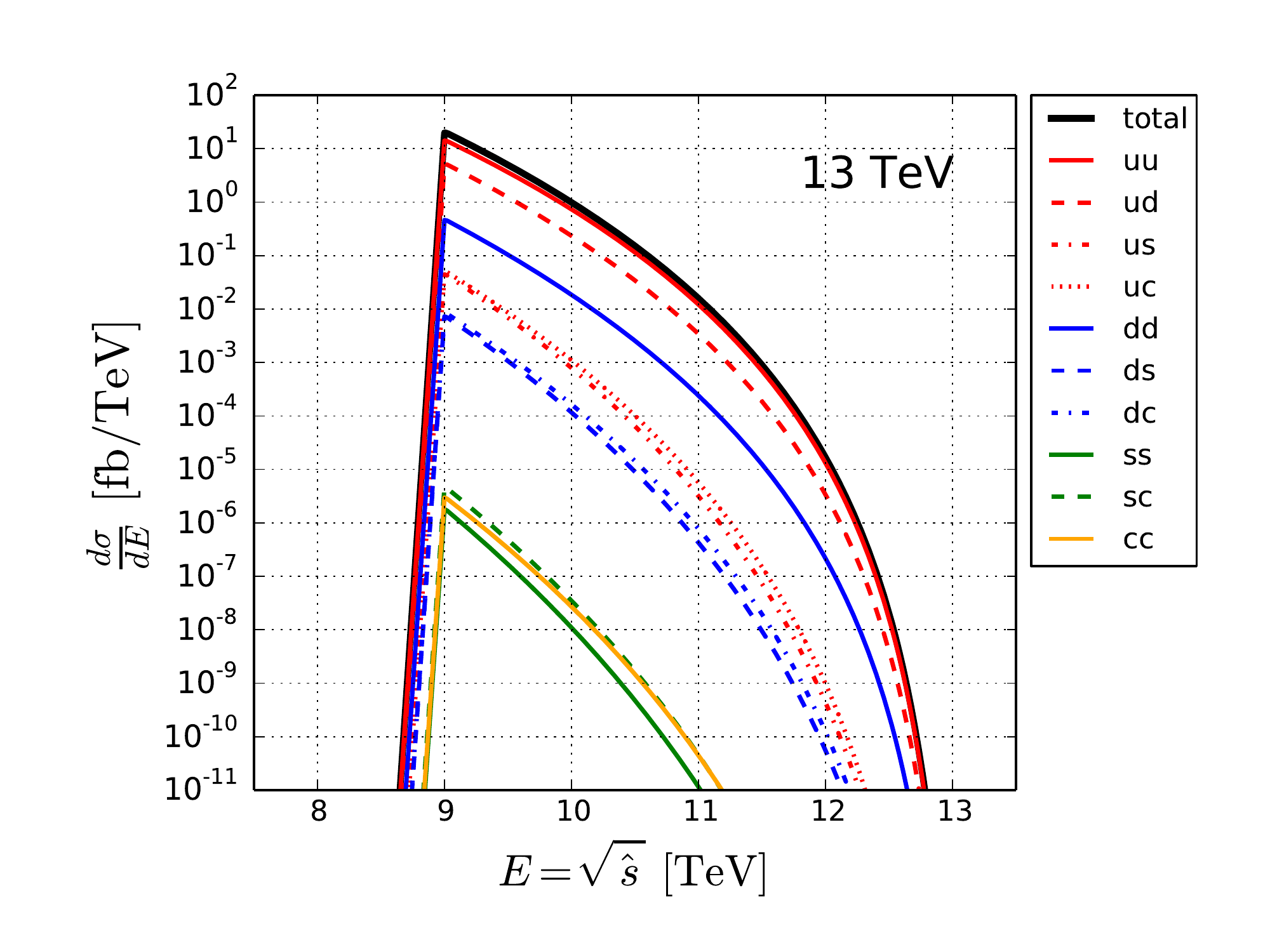} 
\hspace{-0.58cm}
\includegraphics[height=4.28cm]{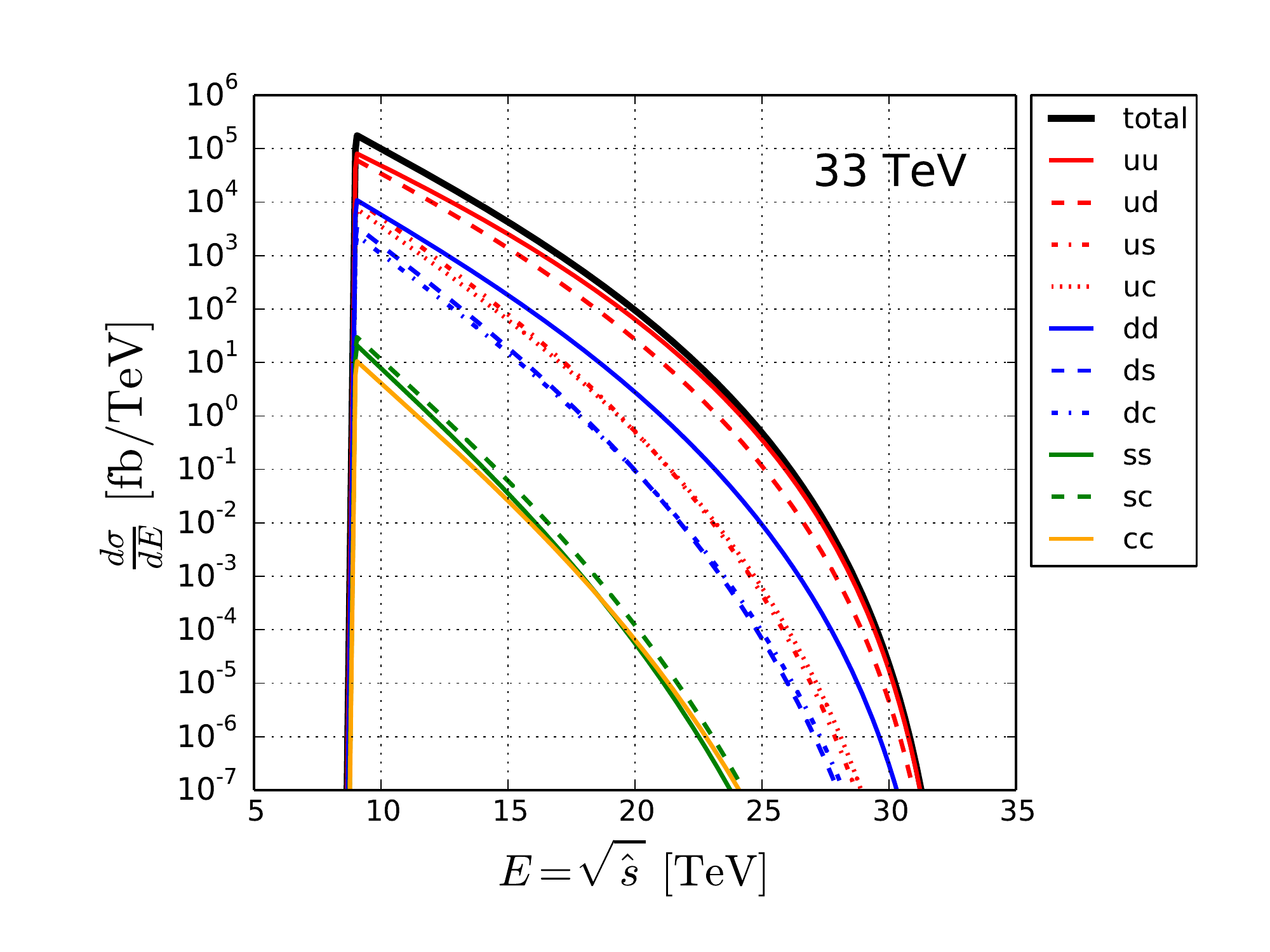} 
\hspace{-0.58cm}
\includegraphics[height=4.28cm]{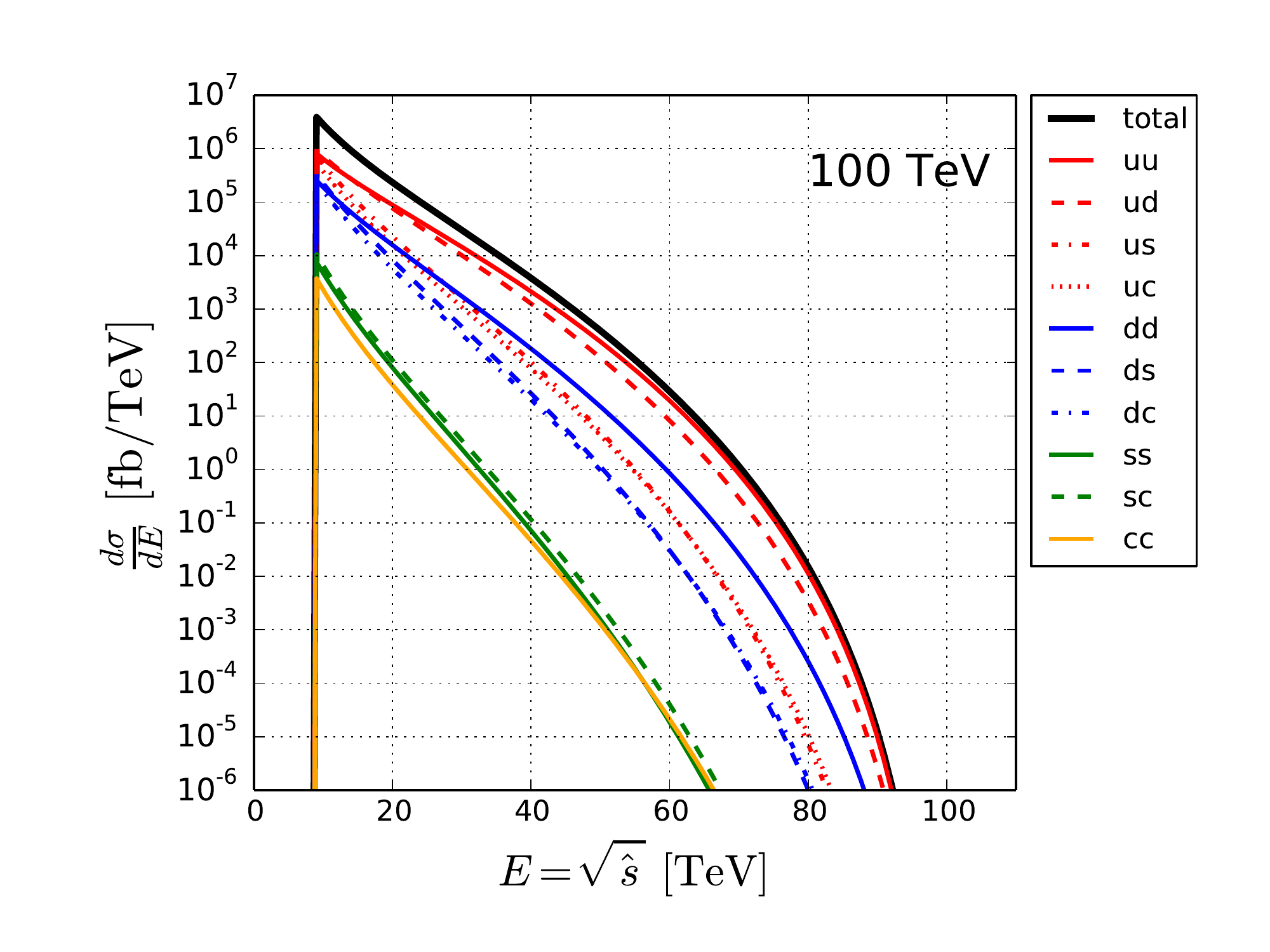} \\
\vspace{-0.5cm}
\end{center}   
\caption{\label{fig:flavours}\it 
Upper panel: Contributions to the cross section for sphaleron transitions from the collisions of different flavours of quarks, 
for the nominal case $E_{\rm CM} = 14$~TeV, $E_{\rm Sph} = 9$~TeV, $c = 2$ and $p = 1$ in (\protect\ref{sigma}) with
$S$ given by (\protect\ref{numericalS}). The contributions of different parton-parton collision processes are colour-coded
as indicated. Lower panels: As above, for the cases $E_{\rm CM} = 13, 33$ and $100$~TeV.
}
\end{figure}

The upper panel of Fig.~\ref{fig:flavours} displays the relative contributions of the collisions
of different quark flavours for the nominal case $E_{\rm CM} = 14$~TeV, $E_{\rm Sph} = 9$~TeV, $c = 2$,
with the normalization corresponding to $p = 1$ in (\ref{sigma}). We see that,
as expected, the dominant contribution to the sphaleron cross section is due to $uu$ collisions, with $ud$ collisions being the
second most important, and other processes contributing $< 3$\% of the total.
Sphaleron production by collisions involving $d$ quarks are suppressed at 14~TeV
because the $u$ parton distribution function is much larger than
that for the $d$ quark at large momentum fraction $x$.

The lower panels of Fig.~\ref{fig:flavours} display the corresponding relative contributions of
different quark flavours for $E_{\rm CM} = 13, 33$ and 100~TeV. As could be expected, the relative contributions
at 13~TeV are quite similar to those at 14~TeV, but the dominance of $uu$ and $ud$ collisions is challenged
increasingly  at 33 and 100~TeV by transitions induced by $dd$ collisions and processes involving
second-generation quarks.

\begin{figure}[t!]
\begin{center}
\includegraphics[height=10cm]{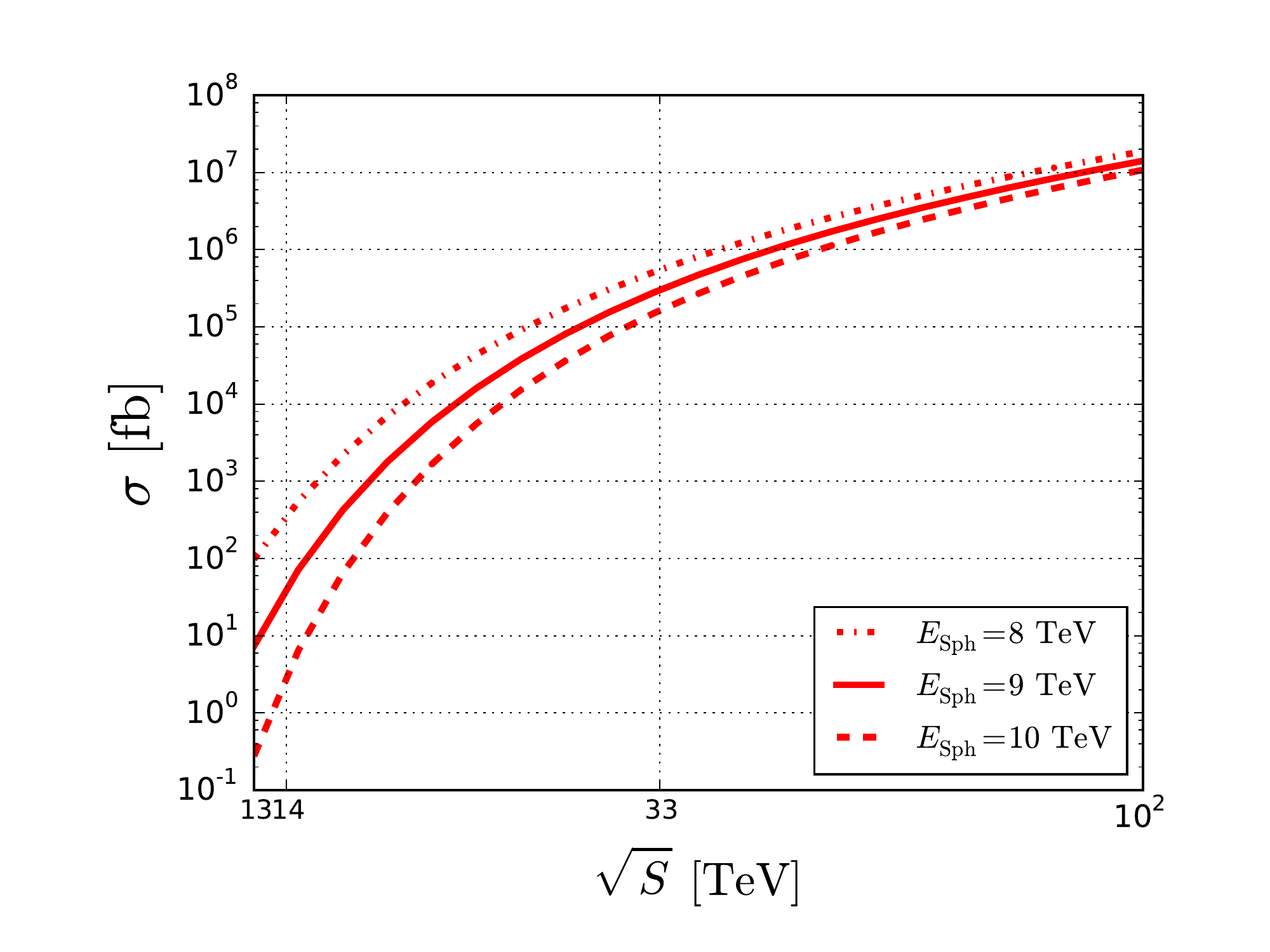}
\end{center}   
\vspace{-0.5cm}
\caption{\label{fig:Edep}\it 
The energy dependence of the total cross section for sphaleron transitions for the nominal
choices $E_{\rm Sph} = 9$~TeV, $c = 2$ and $p = 1$ in (\protect\ref{sigma}) with
$S$ given by (\protect\ref{numericalS}) (solid curve), and for the outlying choices $E_{\rm Sph} = 8$
and $10$~TeV (dot-dashed and dashed lines, respectively. The variations in the curves for $1 \le c \le 4$
are within the widths of the lines. We recall that the overall normalization factor $p$ is quite uncertain.
}
\end{figure}

Our calculations of the $E_{\rm CM}$ dependence of the sphaleron cross section are displayed
in Fig.~\ref{fig:Edep}. The solid curve is for the nominal choices $E_{\rm Sph} = 9$~TeV, $c = 2$ and $p = 1$.
Results for values of the parameter $c \in [1, 4]$ in (\ref{sigma}) are indistinguishable, as could have been anticipated
from the steep fall-off of the sphaleron cross section for $E < E_{\rm Sph}$ seen in Fig.~\ref{fig:flavours}.
We see that the cross section is substantial already at LHC energies {\it if $p = 1$}: specifically we calculate
$\sigma = 7.3$~fb for $E_{\rm CM} = 13$~TeV and $\sigma = 41$~fb for $E_{\rm CM} = 14$~TeV.
However, the value of $p$ must be regarded as {\it very uncertain}, and
these numbers would be reduced substantially if the unknown parameter $p <\ll1$, a
possibility that could only be countered (partially) by increasing $E_{\rm CM}$. The steep rise in the cross section
by a factor $\sim 6$ between $E_{\rm CM} = 13$ and 14~TeV is largely independent of $c$ and $p$,
and puts a premium on developing and
exploiting the full energy capability of the LHC. However, the size of the cross section for sphaleron-induced
processes for the nominal choice $p = 1$ suggests that it would be interesting to look for them during LHC Run~2,
even with just a few fb$^{-1}$ of luminosity at 13~TeV as already accumulated.

The dot-dashed and dashed curves in Fig.~\ref{fig:Edep} are for the cases $E_{\rm Sph} = 8$ and 10~TeV,
which lie far outside the uncertainty in $E_{\rm Sph} \sim 1$\% quoted in~\cite{TW}. It is clear that
the LHC cross section is smaller for larger $E_{\rm Sph}$, and the energy dependence is steeper,
whereas the opposite statements hold for smaller $E_{\rm Sph}$. However, whereas in the former
case sphaleron-induced processes could be more visible in Run~2 of the LHC, even in the latter case
increasing $E_{\rm CM}$ should be a priority for the LHC.

Looking beyond the LHC, Fig.~\ref{fig:Edep} shows that the sphaleron transition rate would
increase significantly at colliders with higher $E_{\rm CM}$. Specifically, for our nominal choices
$E_{\rm Sph} = 9$~TeV, $c = 2$ and $p = 1$ we find sphaleron cross sections
$0.3 ~(141) \times 10^6$ fb at $E_{\rm CM} = 33 ~(100)$~TeV. These can be compared with
the expected $gg \to H$ cross sections at these centre-of-mass energies, which are 
$0.18 ~(0.74) \times 10^6$~fb
at 33 (100) TeV. If these estimates are in the right ball-park, such higher-energy colliders would be
veritable sphaleron factories. However, we emphasize again that the overall magnitude of the
sphaleron transition rate is {\it very uncertain}. One should, perhaps, instead regard Fig.~\ref{fig:Edep}
as showing that higher-energy collisions may provide sensitivity to sphaleron transitions for $p \ll 1$.

\section{Simulations of Sphaleron-Induced Processes}

We turn now to the prospective observability of sphaleron-induced processes, the simplest possibility being
$\Delta n = -1$ processes that give rise to effective interactions involving one member of each electroweak doublet, i.e.,
$e/\nu_e$, $\mu/\nu_\mu$, $\tau/\nu_\tau$, and 3 colours of $u/d$, $c/s$ and $t/b$, leading to transitions of the form
\begin{equation}
q q \; \to \; \bar \ell \; \bar \ell \; \bar \ell \; \bar q \; \bar q \; \bar q \; \bar q \; \bar q \; \bar q \; \bar q \, . 
\label{10}
\end{equation}
{\it A priori}, the leading-order sphaleron-induced processes do not involve electroweak bosons.\footnote{
  There are suggestions that the baryon and lepton number violating processes are enhanced if fermions are produced associated with many
  ${\cal O}(1/\alpha_W)$ electroweak bosons
  \cite{Arnold:1987zg, Ringwald:1989ee, Espinosa:1989qn, McLerran:1989ab, Gibbs:1994cw}.
  We leave the investigation of this possibility for future work.
}
Since the dominant processes are induced by $uu$ and $ud$ collisions: the final states
should contain a single ${\bar u}/{\bar d}$ antiquark, one antilepton from each generation,
three ${\bar c}/{\bar s}$ antiquarks and three ${\bar t}/{\bar b}$ antiquarks, for a total of 10
final state particles.
The initial and final states are constrained so that the total electric charge is conserved.
We simulate the momenta of final state particles according to the phase space.
We also simulate the decays of heavy particles ($t$, $W$ and $\tau$).
We accept only particles with $p_T > 20$ and $|\eta| < 2.5$.
Neutrinos are removed from the list of observable particles.

\begin{figure}[hbt]
\begin{center}
\includegraphics[height=6cm]{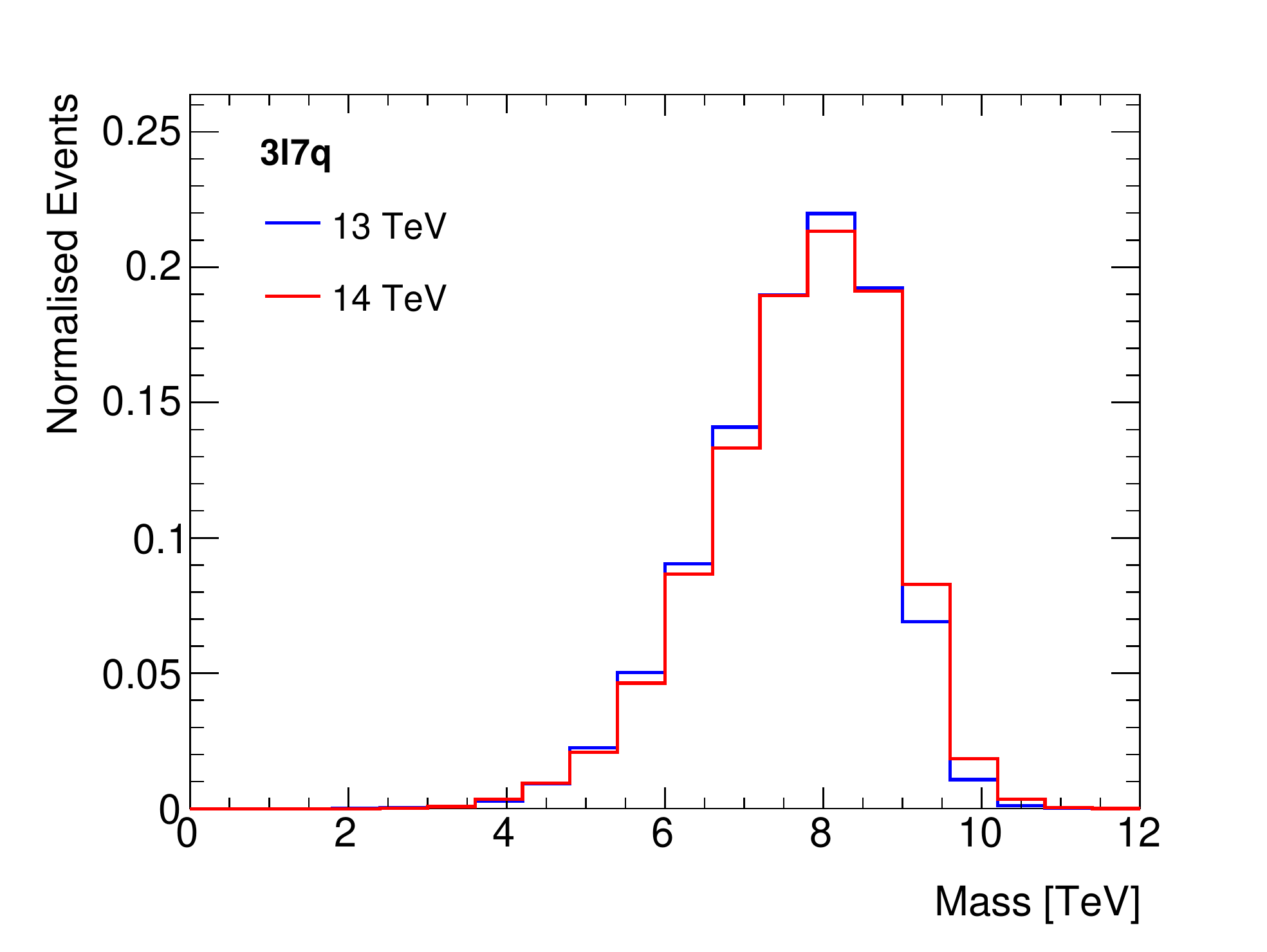}
\includegraphics[height=6cm]{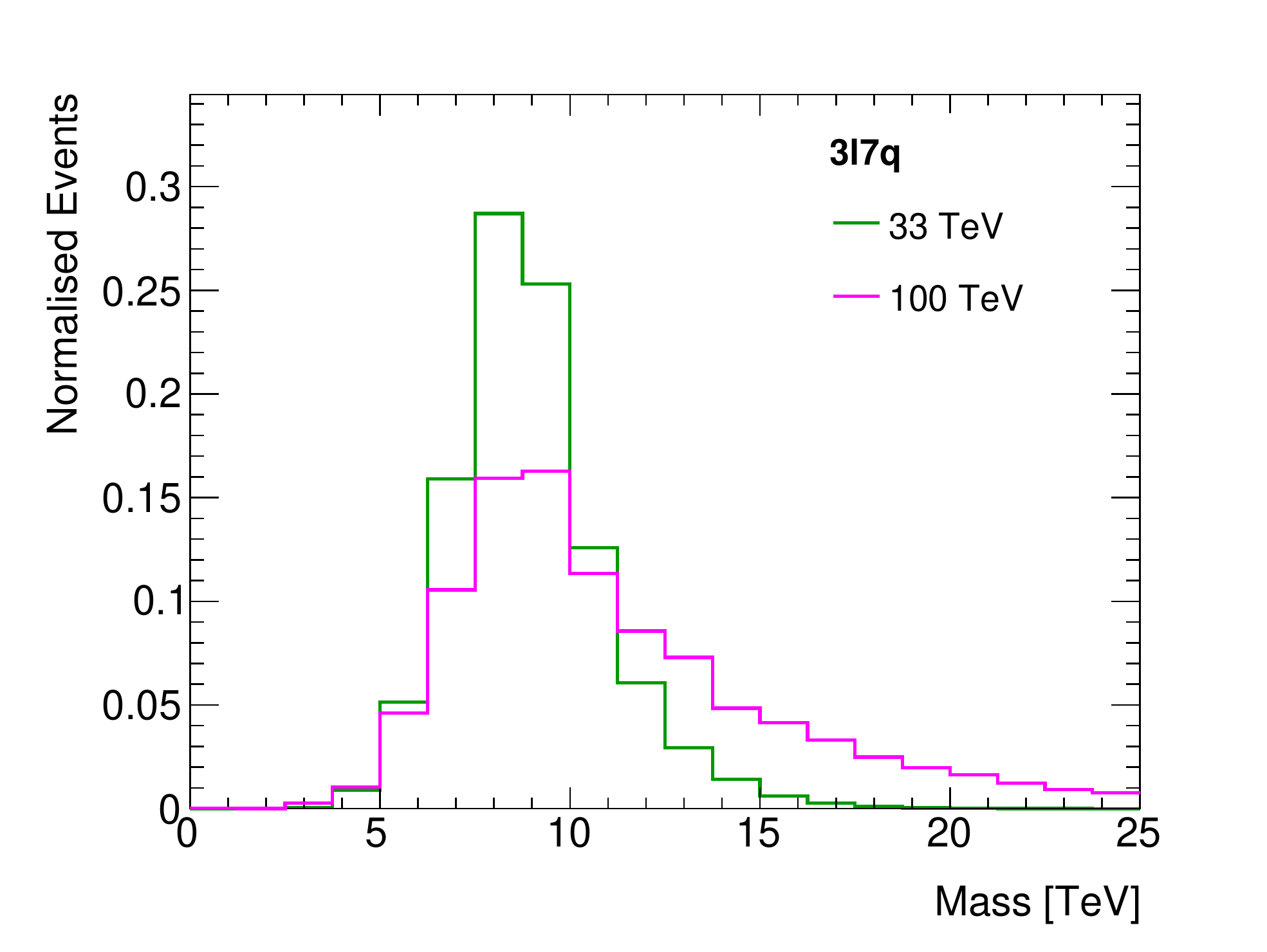} 
\vspace{-0.5cm}
\end{center}   
\caption{\label{fig:masses}\it 
Left panel: Normalized invariant-mass distributions for the observable final-state particles
in sphaleron-induced transitions in LHC collisions at 13 and 14 TeV (blue and red histograms, respectively).
Right panel: Corresponding invariant-mass distributions for future colliders 
at 33 and 100 TeV (green and pink histograms, respectively).
These distributions are calculated for our nominal choices $E_{\rm Sph} = 9$~TeV, $c = 2$ and $p = 1$.
}
\end{figure}

The normalized invariant-mass distributions for the observable final-state particles are shown in Fig.~\ref{fig:masses},
for LHC collisions at 13 and 14 TeV (left panel, blue and red histograms, respectively) and for future colliders 
at 33 and 100 TeV (right panel, green and pink histograms, respectively),
for our nominal choices $E_{\rm Sph} = 9$~TeV, $c = 2$ and $p = 1$. 
As seen in the left panel, the invariant-mass distributions
for 13 and 14 TeV are quite similar, both being peaked at $\sim 8$~TeV and with tails extending to lower masses,
corresponding to events with (multiple) neutrino emission. As seen in the right panel, the corresponding distributions for collisions
at 33 and (particularly) 100 TeV extend to much larger invariant masses.

\begin{figure}[hbt]
\begin{center}
\includegraphics[height=6cm]{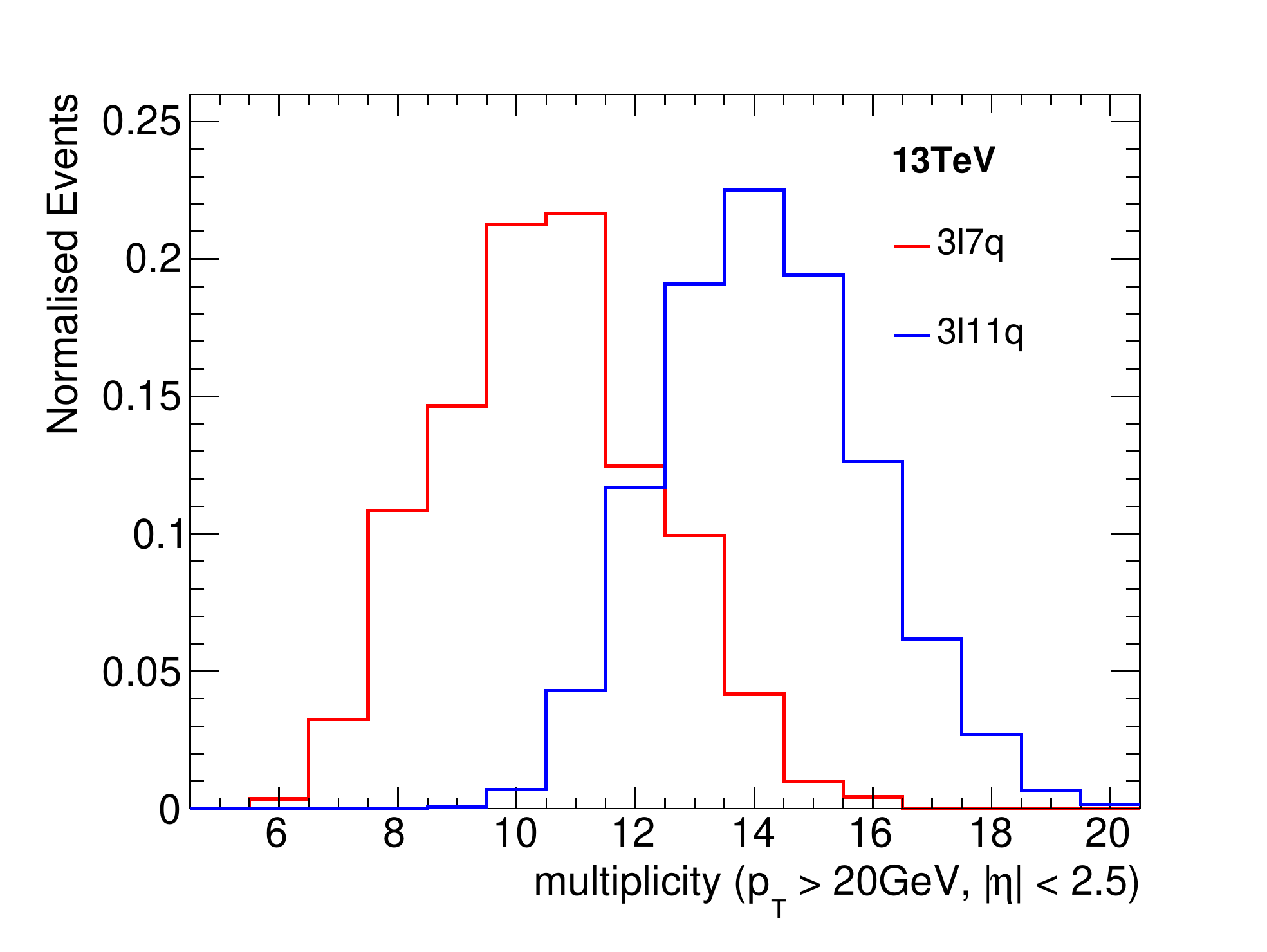}
\includegraphics[height=6cm]{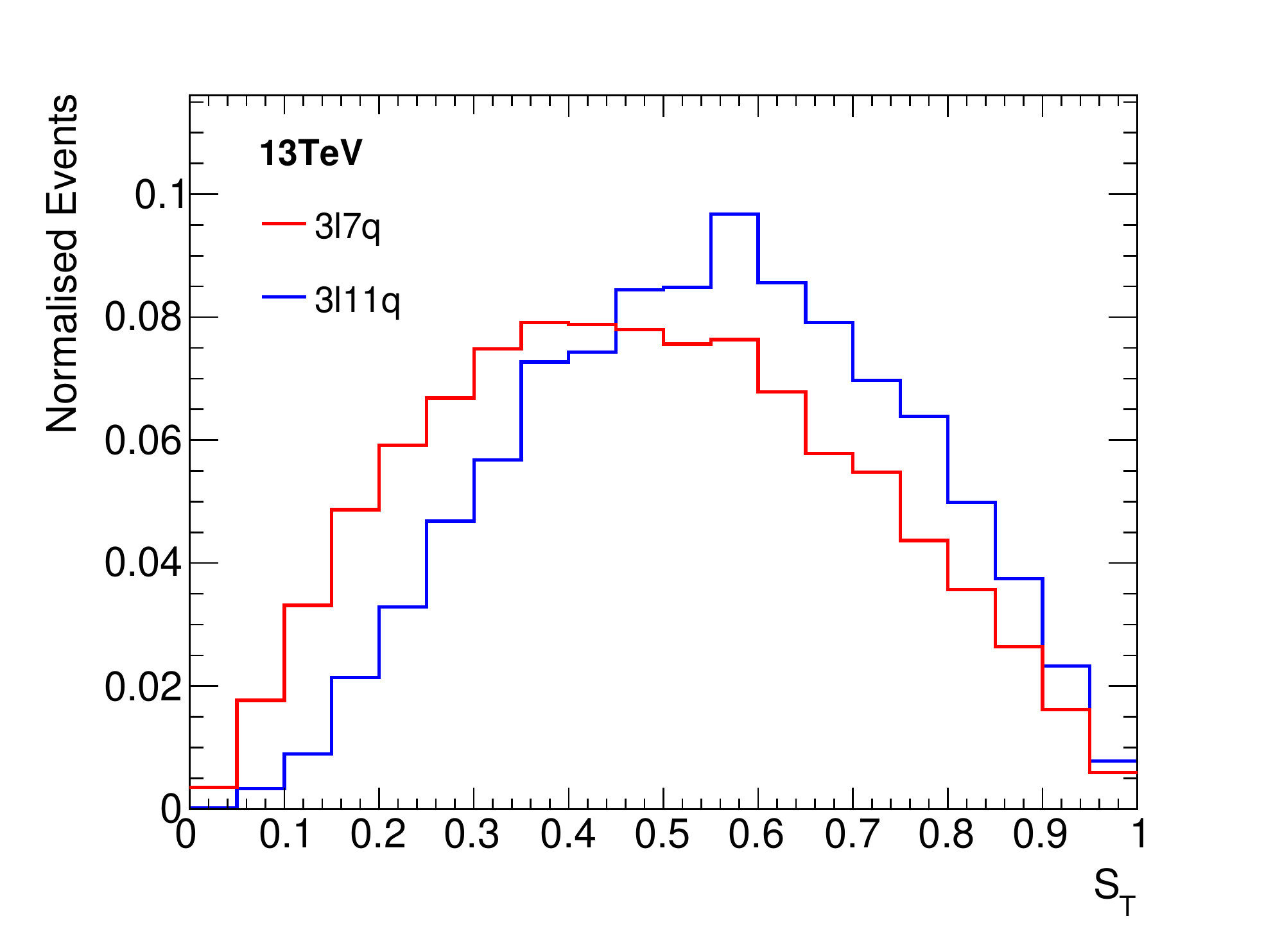} \\
\includegraphics[height=6cm]{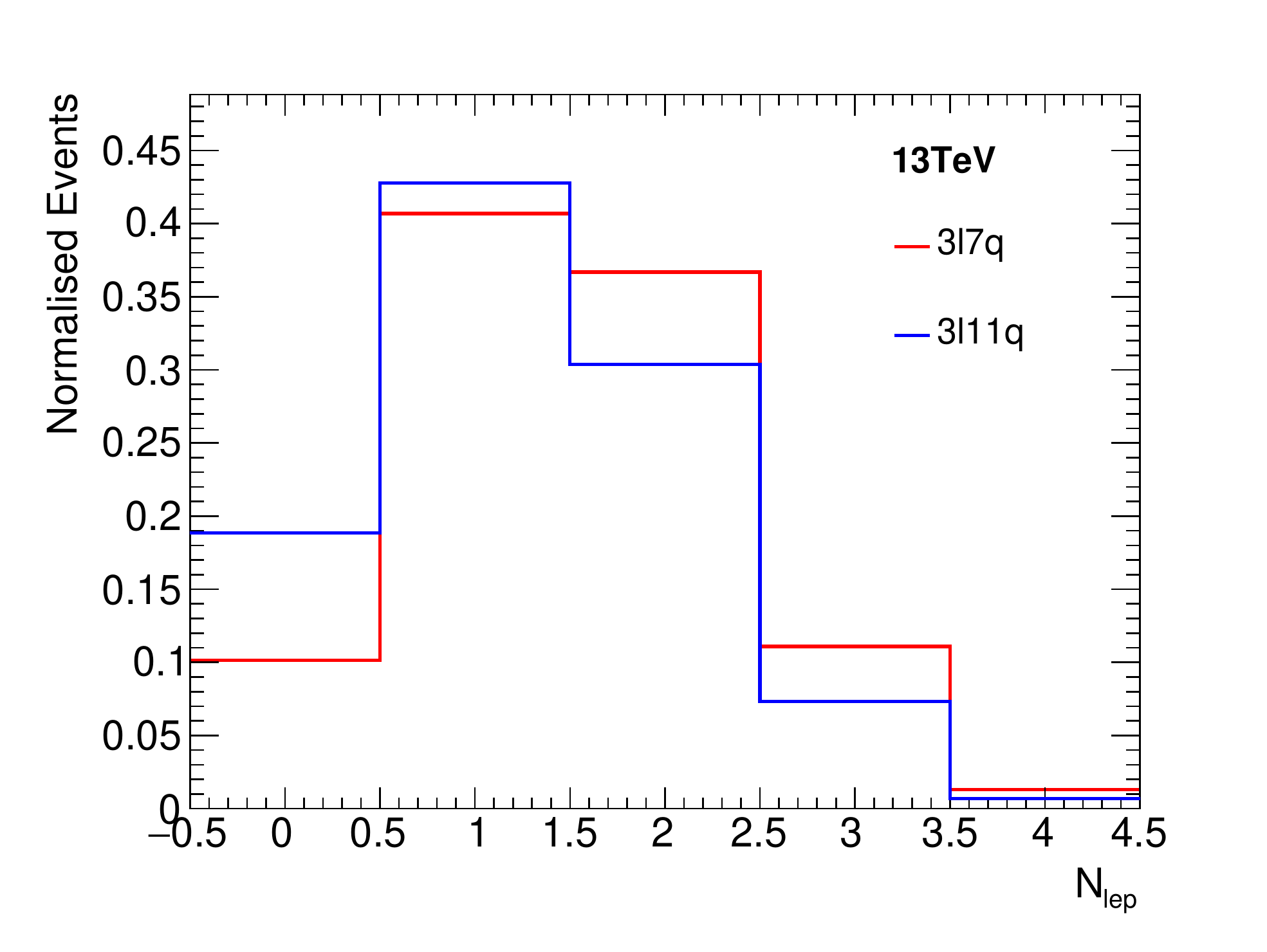}
\includegraphics[height=6cm]{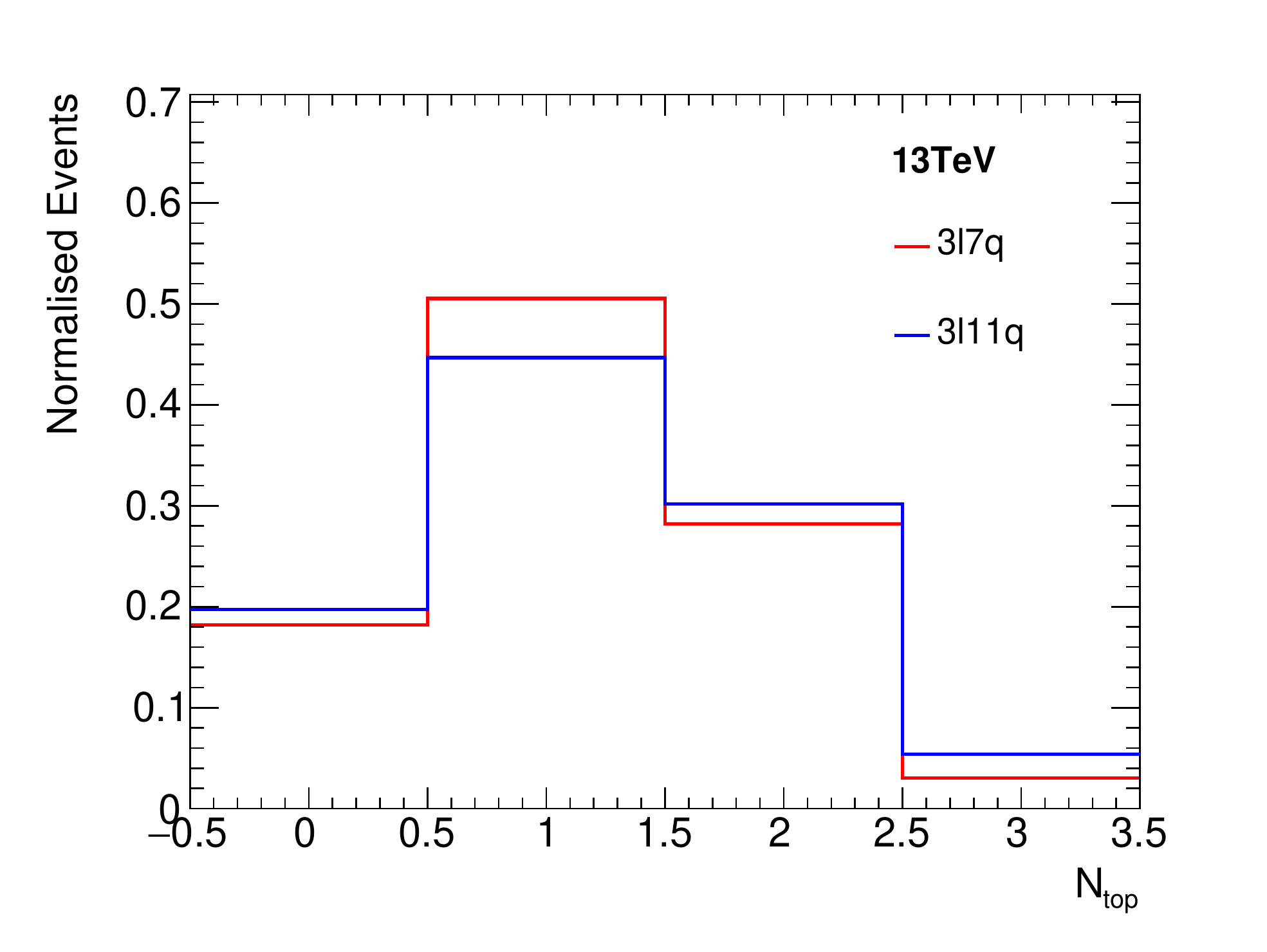} 
\end{center} 
\vspace{-0.5cm}  
\caption{\label{fig:finalstates}\it 
Left panel: Normalized invariant-mass distributions for the observable final state particles
in sphaleron-induced transitions in LHC collisions at 13 and 14 TeV (blue and red histograms, respectively).
Right panel: Corresponding invariant-mass distributions for future colliders 
at 33 and 100 TeV (green and pink histograms, respectively).
These distributions are also calculated for our nominal choices $E_{\rm Sph} = 9$~TeV, $c = 2$ and $p = 1$.
}
\end{figure}

Fig.~\ref{fig:finalstates} displays some more properties of the final states in sphaleron-induced transitions:
the red histograms are for the $\Delta n = -1$ processes leading to 10-particle final states discussed above (\ref{10}).
On should also consider processes with other values of $\Delta n$, the next simplest being the $\Delta n = +1$ process
that leads to 14-particle final states:
\begin{equation}
q \; q \; \to \; \ell \; \ell \; \ell \; q \; q \; q \; q \; q \; q \; q \; q \; q \; q \; q \, ,
\label{14}
\end{equation}
whose simulation yields the blue histograms in Fig.~\ref{fig:finalstates}.
The difference between these nominal multiplicities is visible in the upper left panel.
The multiplicity may exceed the nominal value if gluon radiation or other higher-order QCD processes
yield additional final-state partons satisfying our chose acceptance cuts: $p_T > 20$~GeV and $| \eta | < 2.5$.
On the other hand, the visible multiplicity may be reduced if some final-state particles fail these acceptance
cuts and/or if there are neutrinos in the final state.

The upper right panel of Fig.~\ref{fig:finalstates} shows the distribution in the sphericity, $S_T$, for sphaleron-induced 
final states with 10 and 14 final-state particles as red and blue histograms, respectively. Both distributions are relatively
broad, being peaked at $S_T \sim 0.4$ and 0.6, respectively.
The lower left panel of Fig.~\ref{fig:finalstates} displays histograms of the number of charged leptons, $N_{\rm lep} = N_e + N_\mu$,
satisfying the nominal acceptance cuts. As expected, $N_{\rm lep} \le 3$, with smaller numbers of charged leptons
in events with final state neutrinos and/or charged leptons outside the nominal acceptance range. Finally, the lower right
panel of Fig.~\ref{fig:finalstates} shows histograms of the numbers of top quarks in the sphaleron-induced final states.
The most common outcome is to observe just one top quark, followed by final states with two top quarks.
There are relatively few final states with no top quarks, and even fewer with three top quarks. The difference between
the cases with one and two top quarks is due to the phase-space suppression of final states with more top quarks than bottom quarks,
as is that between the cases with zero and three top quarks.  

\begin{figure}[!t]
\begin{center}
\includegraphics[height=6cm]{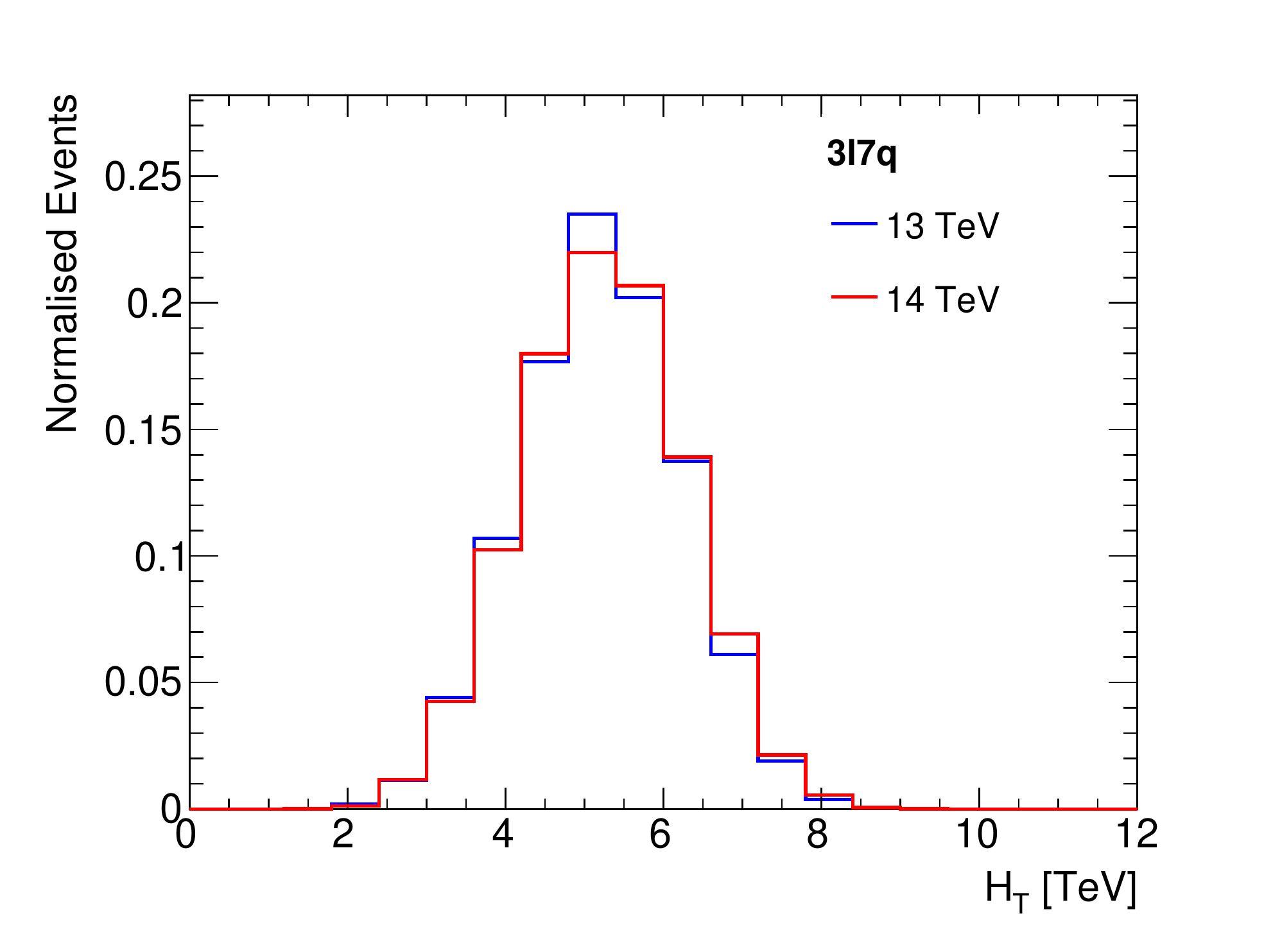}
\includegraphics[height=6cm]{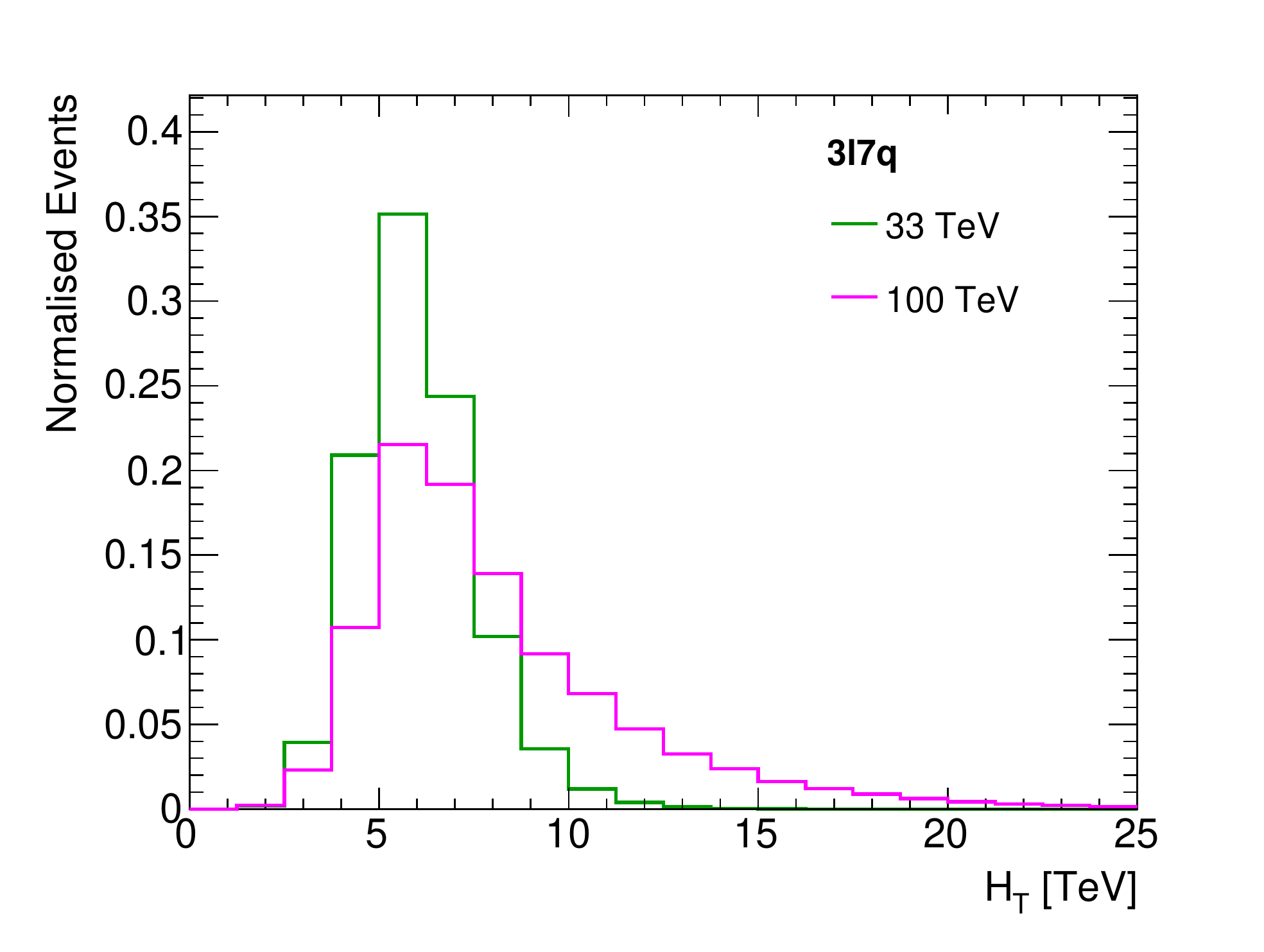}
\includegraphics[height=6cm]{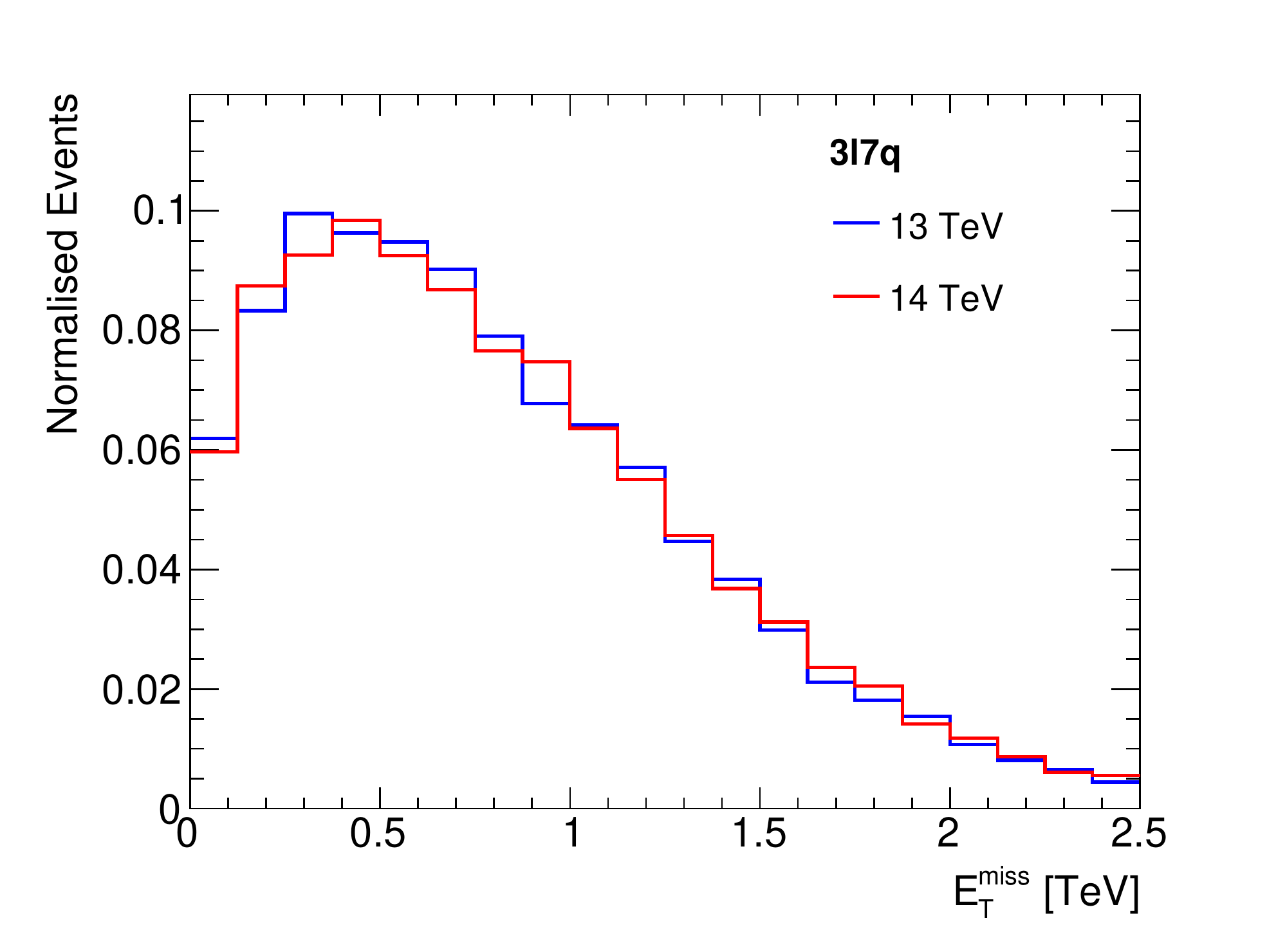}
\includegraphics[height=6cm]{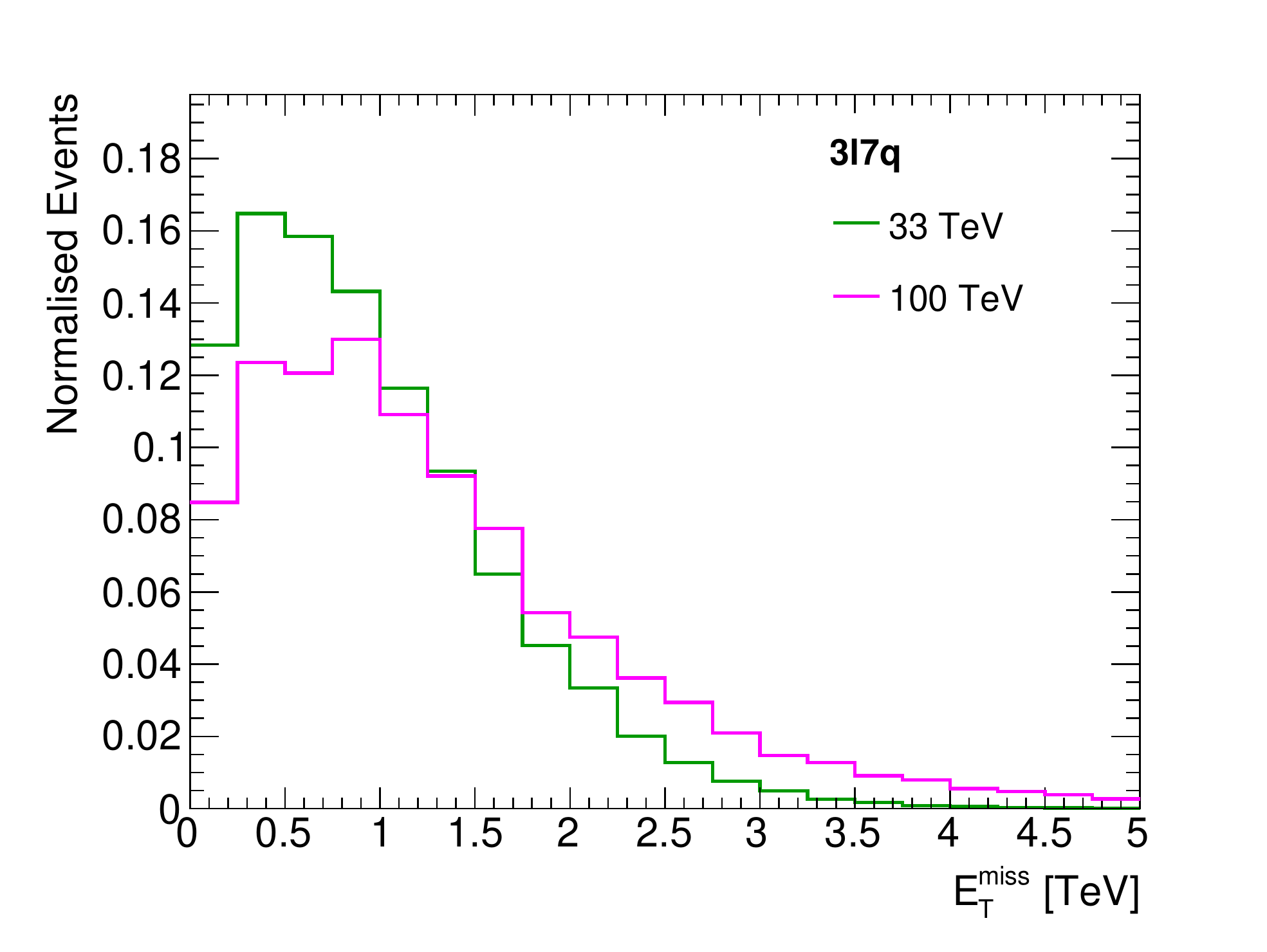}
\end{center} 
\vspace{-0.5cm}  
\caption{\label{fig:pTsummet}\it 
Upper panels: Normalized distributions in $H_T \equiv \sum p_T^{\rm jet}$ for the observable final state particles
in sphaleron-induced transitions in LHC collisions at 13 and 14 TeV in the left panel and at 33 and 100 TeV
in the right panel. Lower panels: Normalized distributions in $E_T^{\rm miss}$ for the two LHC energies
in the left panel and the two future collider energies in the right panel. Again,
these distributions are calculated for our nominal choices $E_{\rm Sph} = 9$~TeV, $c = 2$ and $p = 1$.
}
\end{figure}

Additional properties of 10-particle sphaleron final states are shown in Fig.~\ref{fig:pTsummet}, where we compare
distributions at different LHC energies, 13 and 14~TeV, in the left panels (blue and red histograms, respectively),
and possible future collider energies, 33 and 100~TeV, in the right panels (green and pink histograms, respectively).
The upper panels display the distributions in $H_T \equiv \sum p_T^{\rm jet}$ and the lower panels the distributions in $E_T^{\rm miss}$,
where the former is used in the ATLAS microscopic black hole search~\cite{ATLAS}.
The $H_T$ distributions at the two LHC energies are very similar, both being peaked at $\sim 6$~TeV. 
The distributions at the two future collider energies peak at somewhat higher energies $\sim 7$~TeV, but with longer tails at higher values of $H_T$, particularly at 100~TeV. 
The $E_T^{\rm miss}$ distributions at the two LHC energies are also very similar, both being peaked at $\sim 0.5$~TeV. The distributions at 33 and 100 TeV are also peaked at $\sim 0.7$~TeV, but with longer tails to higher values, particularly at 100~TeV.

\section{Analysis of ATLAS 2015 Data}

The ATLAS Collaboration has recently published the (null) results of a search for microscopic
black holes using $\sim 3$~fb$^{-1}$ of data at 13~TeV recorded in 2015~\cite{ATLAS}. This analysis was based on
measurements of the numbers of events in search regions (SRn$_{\rm jet}$) defined by cuts in the number of jets, 
$n_{\rm jet} \ge 3$ to $8$, accompanied by cuts in
$H_T \gtrsim 5$~TeV. We now compare the ATLAS measurements with our simulations of the final states induced
by sphaleron transitions. 

\begin{figure}[!t]
\begin{center}
\includegraphics[height=9cm]{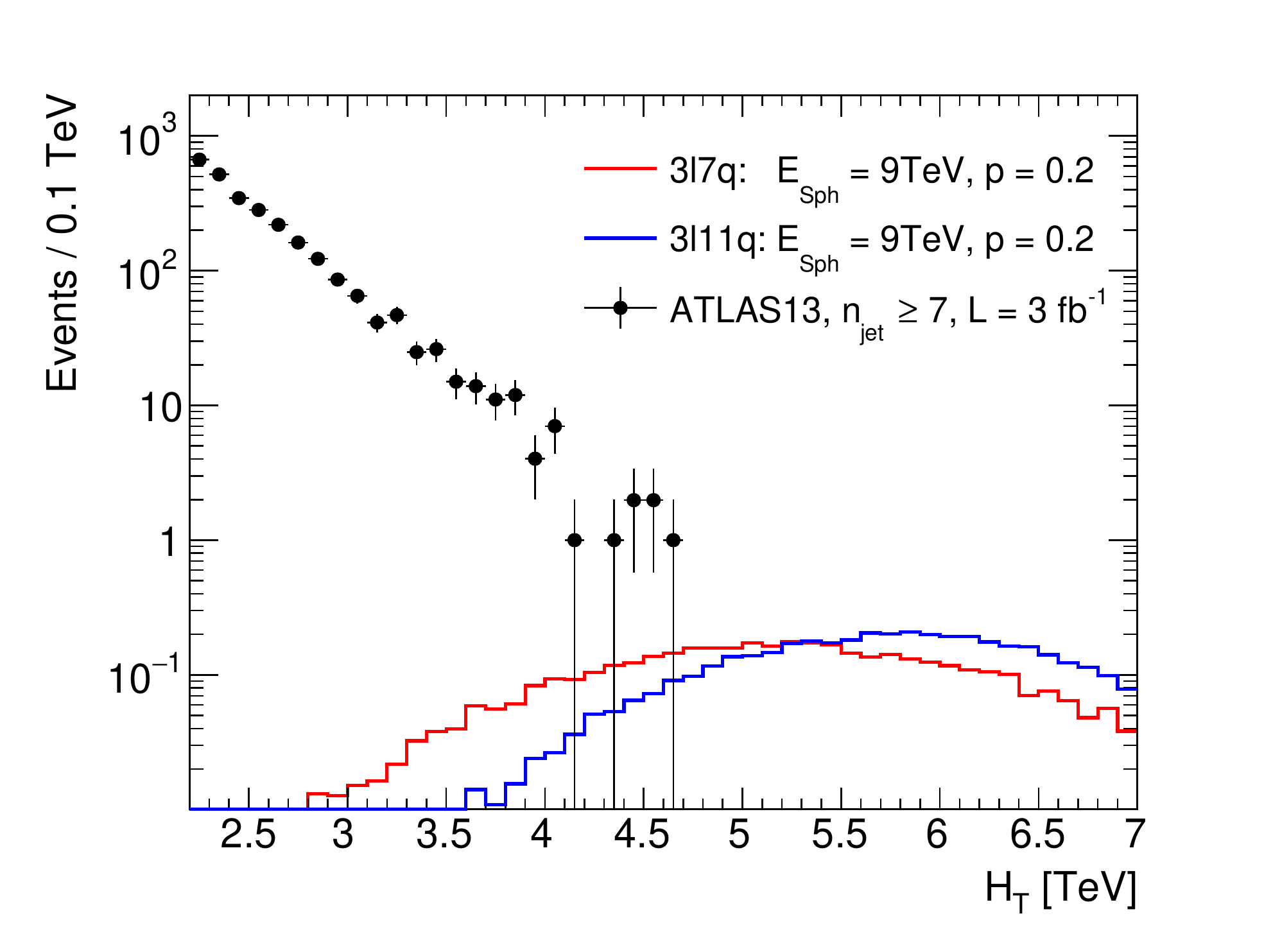}
\end{center} 
\vspace{-0.5cm}  
\caption{\label{fig:ATLAS}\it 
Comparison of the numbers of events with $n_{\rm jet} \ge 3$ measured by ATLAS in
$\sim 3$/fb of data at 13~TeV in bins of $H_T$, compared with simulations for
$E_{\rm Sph} = 9$~TeV and $c = 2$ of $\Delta n = -1$ sphaleron transitions to final states
with 3 antileptons and 7 antiquarks (red histogram) and $\Delta n = +1$ transitions to final
states with 3 leptons and 11 quarks (blue histogram). 
}
\vspace{0.2cm}
\end{figure}

\begin{figure}[t!]
\begin{center}
\includegraphics[height=6cm]{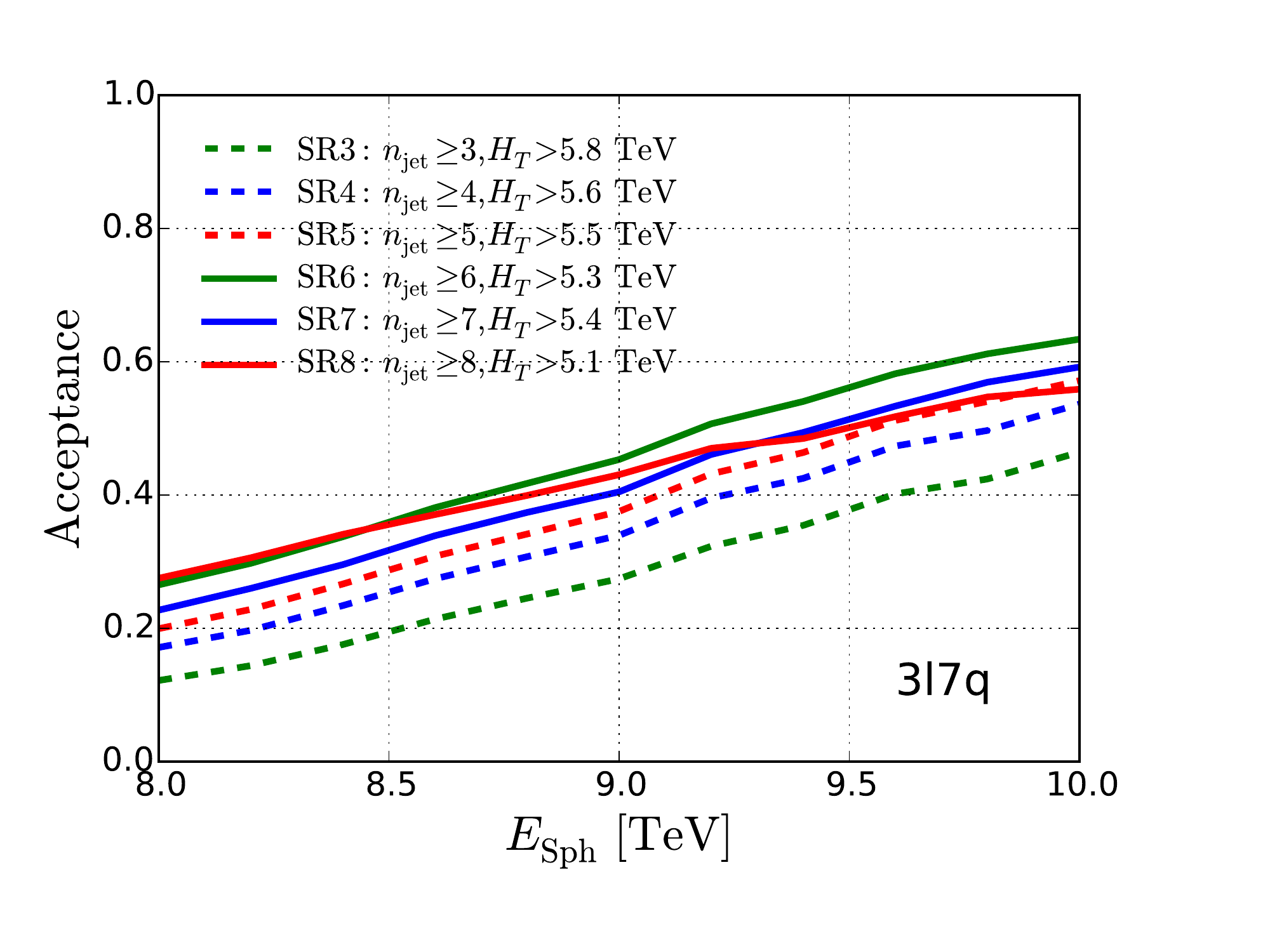}
\includegraphics[height=6cm]{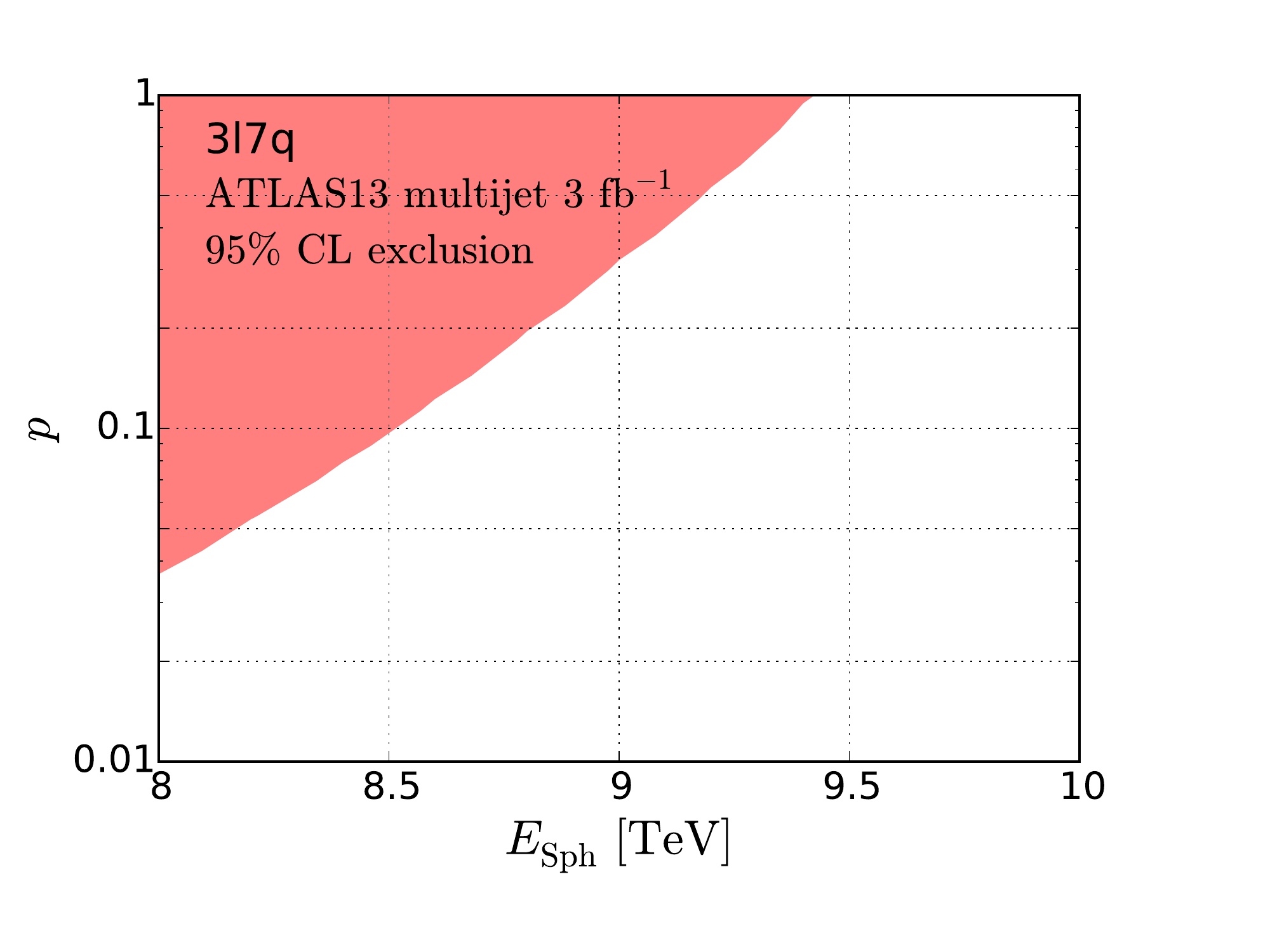}
\end{center}   
\vspace{-0.5cm}
\caption{\label{fig:exclusion}\it 
Right panel: Acceptances for sphaleron-induced $\Delta n = - 1$ transitions
in ATLAS event selections with different cuts in $(n_{\rm jet}, H_T)$, as functions of $E_{\rm Sph}$.
Left panel: The exclusion in the $(E_{\rm Sph}, p)$ plane of $\Delta n = - 1$ transitions obtained by recasting the ATLAS 2015 search
for microscopic black holes using $\sim 3$/fb of data at 13~TeV. The variation in the exclusion for $1 \le c \le 4$ is negligible.
}
\end{figure}

Fig.~\ref{fig:ATLAS} compares the ATLAS
measurements for $n_{\rm jet} \ge 7$ in bins of $H_T \le 7$~TeV with sphaleron
simulations for $E_{\rm Sph} = 9$~TeV and $p = 0.2$ (the results are insensitive to $c$).
We see that events due to sphaleron transitions are expected to have a broad distribution in $H_T$,
with a large fraction having $H_T \gtrsim 5$~TeV. We focus initially on the case of $\Delta n = -1$ 
transitions, which yield final states with 3 antilepton + 7 antiquarks (\ref{10}), corresponding to the red histogram
in Fig.~\ref{fig:ATLAS}. The corresponding values of the acceptances for these final states
in the different ATLAS search regions SR3, ..., SR8 as functions of the sphaleron barrier height
$E_{\rm Sph} \in [8, 10]$~TeV are shown in the left panel of Fig.~\ref{fig:exclusion}. We note that
characteristic values of the acceptances for the nominal $E_{\rm Sph} = 9$~TeV are $\gtrsim 0.4$
for SR6, SR7 and SR8.

We may therefore recast the ATLAS search as a relatively efficient search for $\Delta n = -1$ sphaleron-induced
transitions. For each value of $E_{\rm Sph}$, we select the SRn that is expected to yield the best
limit, finding that SR8 is expected to be the most sensitive for $E_{\rm Sph} \lesssim 9.3$~TeV
whereas SR7 is the most sensitive for $E_{\rm Sph} \gtrsim 9.3$~TeV. 
The exclusion limit resulting from this recasting of the ATLAS black hole search is shown in the right panel of Fig.~\ref{fig:exclusion}. 
We display the 95\% CL constraint in the $(E_{\rm Sph}, p)$ plane, which is quite insensitive to $c \in [1, 4]$.
We note that this preliminary result already excludes $p = 1$ for the nominal value of $E_{\rm Sph} = 9$~TeV.

\begin{figure}[!t]
\begin{center}
\includegraphics[height=6cm]{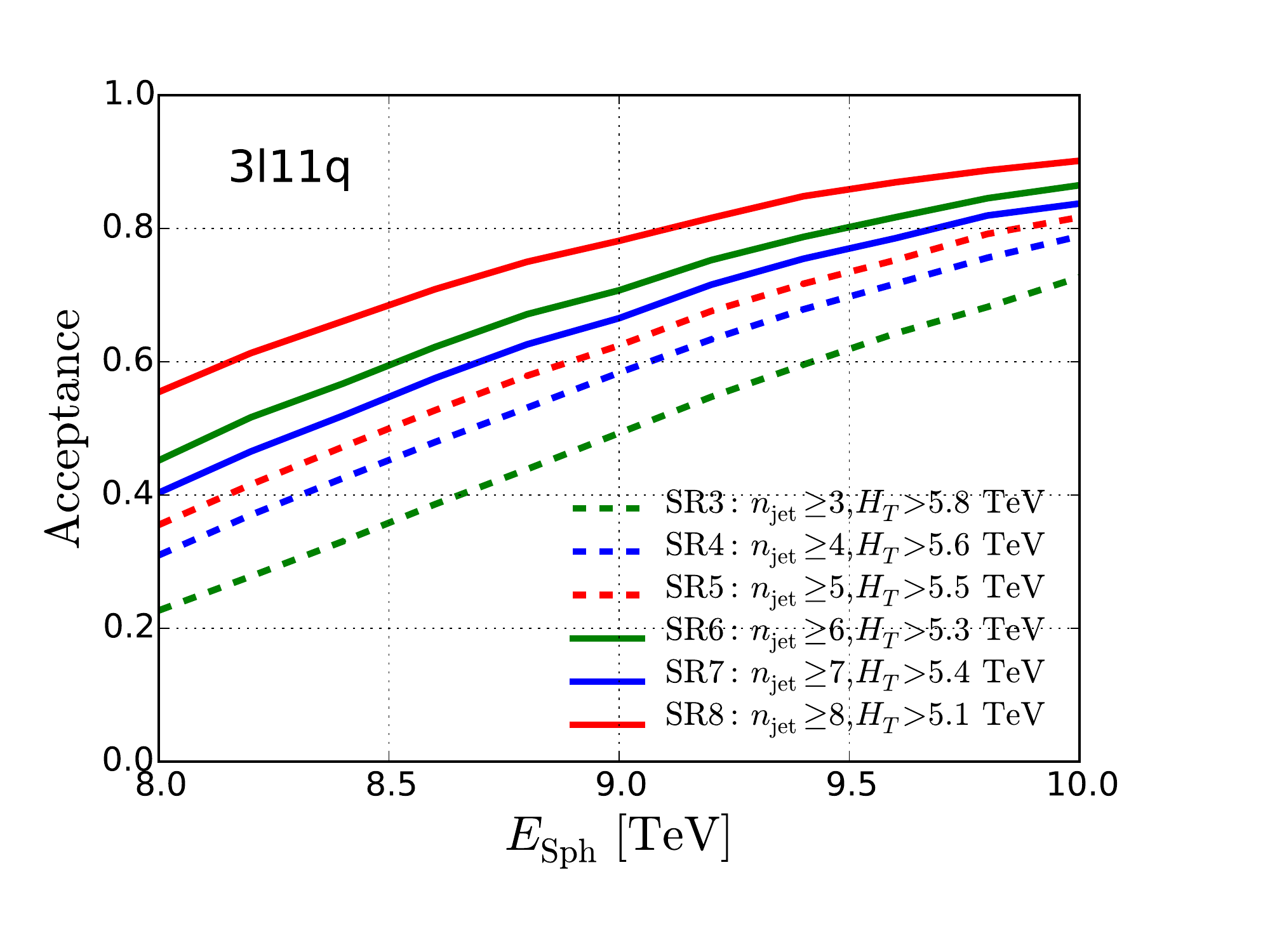}
\includegraphics[height=6cm]{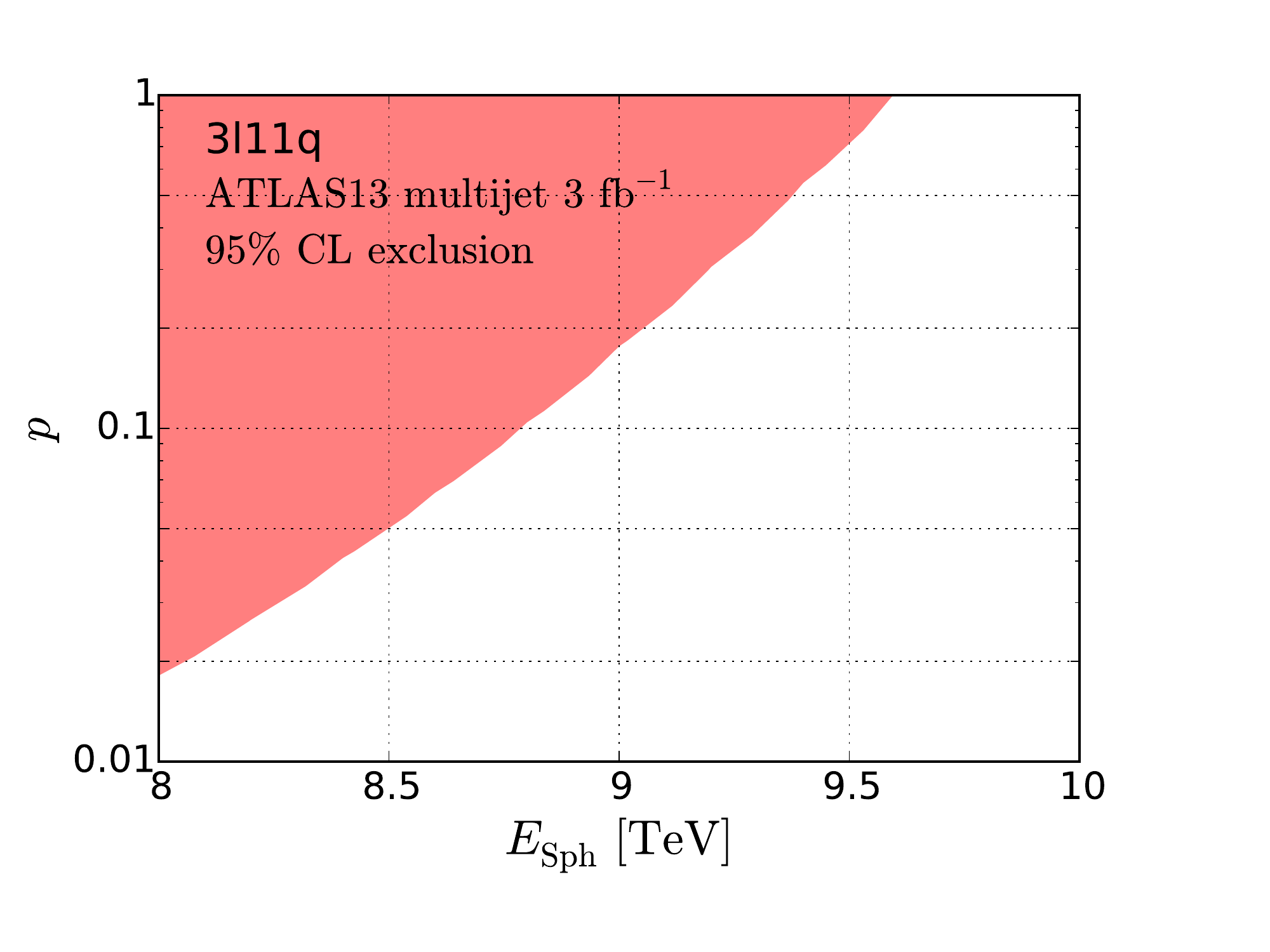}
\end{center} 
\vspace{-0.5cm}  
\caption{\label{fig:ATLAS14}\it 
Left panel: Acceptances as in right panel of Fig.~\protect\ref{fig:ATLAS}, but for $\Delta n = + 1$
sphaleron-induced transitions to 14-particle final states. Right panel: The exclusion in the $(E_{\rm Sph}, p)$ plane, 
as in Fig.~\protect\ref{fig:exclusion} but for sphaleron-induced transitions to 14-particle final states.
}
\end{figure}

Thus far, we have discussed $\Delta n = -1$ sphaleron transitions in which two quarks collide to yield
3 antileptons and 7 antiquarks, and now we consider the next simplest possibility of a $\Delta n = +1$ sphaleron transition
in which two quarks collide to yield 3 leptons and 11 quarks.  The left panel of Fig.~\ref{fig:ATLAS} shows
the simulated $H_T$ distribution for this possibility as a blue histogram, which is shifted to larger values
than for the $\Delta n = -1$ sphaleron transitions. Correspondingly, the acceptances in the ATLAS search
regions are higher for $\Delta n = + 1$ transitions, as seen in the left panel of Fig.~\ref{fig:ATLAS14},
reaching $\sim 0.8$ for SR8 for the nominal $E_{\rm Sph} = 9$~TeV. Consequently, the 95\% CL
exclusion in the $(E_{\rm Sph}, p)$ plane for $\Delta n = + 1$ transitions is correspondingly stronger than for $\Delta n = -1$ transitions,
as seen in the right panel of Fig.~\ref{fig:ATLAS14}, excluding $p \simeq 0.2$ for the nominal $E_{\rm Sph} = 9$~TeV~\footnote{Similarly,
there would be even stronger exclusions for $|\Delta n| > 1$ transitions.}.

\section{Future Prospects}

Run 2 of the LHC is expected to yield $\sim 100$~fb$^{-1}$ of data at 13~TeV, which should enable the
sensitivity to $p$ to be improved to $ \sim 0.01$ for $E_{\rm Sph} = 9$~TeV, which could be
improved with an optimized, targeted analysis of the final states in sphaleron-induced transitions.

For example, as was pointed out in~\cite{TW}, $\Delta n = - 1$ sphaleron-induced processes
would yield final states with multiple positively-charged leptons: $e^+, \mu^+$ and/or $\tau^+$. 
In particular, 1/8 of the final states would contain the distinctive combination of all three
positively-charged leptons: $e^+ + \mu^+ + \tau^+$.
Also, every $\Delta n = - 1$ sphaleron-induced event would contain 0, 1, 2 or 3
top antiquarks accompanied by 3, 2, 1 or 0 bottom antiquarks. 
Therefore, every sphaleron-induced final state should contain multiple bottom antiquarks, produced either directly or in antitop decays.
Assuming the nominal value $E_{\rm Sph} = 9$~TeV,
we have calculated the phase space factors for final states in $\Delta n = - 1$ processes
containing 1, 2 or 3 top antiquarks, which are reduced by 0.90, 0.75 and 0.62 relative to topless final states. 
Including combinatorial factors of 3 for the 1- and 2-top final states
and the constraint of charge conservation and detector acceptance
we found the ratios of 0-, 1-, 2- and 3-top final states to be $1 : 2.83 : 1.56 : 0.17$,
as can be seen from the bottom right panel of Fig.~\ref{fig:finalstates}.
The final states containing top antiquarks may therefore provide distinctive signatures.
Using such antilepton, bottom and top antiquark signatures might improve the Run-2
sensitivity significantly, particularly if both ATLAS and CMS searches could be combined.

The sensitivity could be further improved by a factor $\sim 6$ if the LHC could make
collisions at 14~TeV, and by another factor of 30 with 3000 fb$^{-1}$ of luminosity, pushing the
sensitivity to $p < 10^{-4}$ for $E_{\rm Sph} = 9$~TeV. The sensitivity could be further
improved to $p \sim 10^{-11}$ for two experiments each with 20,000 fb$^{-1}$ of luminosity at 100~TeV
in the centre of mass.
The fact that future searches at the LHC and a possible future collider have such interesting prospective
sensitivities to sphaleron-induced transitions reinforces the importance of assessing the
reliability of the TW estimate of the sphaleron transition rate.
Both the exponential factor $S(E)$ and the pre-exponential
factor $p$ need close scrutiny. Our exploratory study shows that this is not just an academic study,
but could have exciting implications for future $pp$ collider experiments.

\newpage
\section*{Acknowledgements}

The work of JE was supported partly by the London Centre for Terauniverse Studies (LCTS), using funding from the European Research Council via the Advanced Investigator Grant 26732, and partly by the STFC Grant ST/L000326/1. He thanks Henry Tye and Andy Cohen for instructive discussions.

\bibliographystyle{hieeetr}
\bibliography{ES-KS.bib}


\end{document}